\documentclass[twocolumn]{aastex62}

\usepackage{graphicx}
\usepackage{subfigure}
\usepackage{bm}
\usepackage{amsmath}
\usepackage{xspace}

\usepackage{savesym}
\savesymbol{tablenum}
\usepackage{siunitx}
\restoresymbol{SIX}{tablenum}
\usepackage{lineno}

\received{}
\revised{}
\accepted{}

\newcommand{\tto}{\ensuremath{T_\mathrm{21cm}}\xspace}
\newcommand{\meantto}{\ensuremath{\bar{T}_\mathrm{21cm}}\xspace}
\newcommand{\stdto}{\ensuremath{\sigma_\mathrm{21cm}}\xspace}
\newcommand{\skewtto}{\ensuremath{\widetilde{\mu}_{3,\mathrm{21cm}}}\xspace}

\DeclareSIUnit[number-unit-product = {}]\Mpc{Mpc}
\DeclareSIUnit[number-unit-product = {}]\cMpc{cMpc}
\DeclareSIUnit[number-unit-product = {}]\ckpc{ckpc}
\DeclareSIUnit[number-unit-product = {}]\Mpch{\si{\per\h\mega\pc}}
\DeclareSIUnit[number-unit-product = {}]\kpch{\si{\per\h\kilo\pc}}
\DeclareSIUnit[number-unit-product = {}]\cMpch{\si{\per\h\cMpc}}
\DeclareSIUnit[number-unit-product = {}]\ckpch{\si{\per\h\ckpc}}
\DeclareSIUnit[number-unit-product = {}]\iMpch{\si{\h\per\mega\pc}}
\DeclareSIUnit[number-unit-product = {}]\ikpch{\si{\h\per\kilo\pc}}
\DeclareSIUnit[number-unit-product = {}]\icMpch{\si{\h\per\cMpc}}
\DeclareSIUnit[number-unit-product = {}]\ickpch{\si{\h\per\ckpc}}
\DeclareSIUnit[number-unit-product = {}]\pc{pc}
\DeclareSIUnit[number-unit-product = {}]\h{\textit{h}}
\DeclareSIUnit[number-unit-product = {}]\mK{\milli\K}

\shorttitle{}
\shortauthors{}

\begin{document}

\title{Investigating X-ray sources during the epoch of reionization with the 21 cm signal}

\correspondingauthor{Qing-Bo Ma}
\email{maqb@gznu.edu.cn}

\author[0000-0001-9493-4565]{Qing-Bo Ma}
\affil{Guizhou Provincial Key Laboratory of Radio Astronomy and Data Processing, \\
Guizhou Normal University, Guiyang 550001, PR China}
\affil{School of Physics and Electronic Science, Guizhou Normal University, Guiyang 550001, PR China}

\author{Benedetta Ciardi}
\affiliation{Max-Planck-Institut f\"ur Astrophysik, Karl-Schwarzschild-Stra\ss e 1, D-85748 Garching bei M\"unchen, Germany}

\author{Marius B. Eide}
\affiliation{Max-Planck-Institut f\"ur Astrophysik, Karl-Schwarzschild-Stra\ss e 1, D-85748 Garching bei M\"unchen, Germany}

\author{Philipp Busch}
\affiliation{Department of Natural Science, The Open University of Israel, 1 University Road, P. O. Box 808, Raanana 43107, Israel}
\affiliation{Max-Planck-Institut f\"ur Astrophysik, Karl-Schwarzschild-Stra\ss e 1, D-85748 Garching bei M\"unchen, Germany}

\author[0000-0002-1301-3893]{Yi Mao}
\affiliation{Department of Astronomy, Tsinghua University, Beijing 100084, China}

\author{Qi-Jun Zhi}
\affil{School of Physics and Electronic Science, Guizhou Normal University, Guiyang 550001, PR China}
\affil{Guizhou Provincial Key Laboratory of Radio Astronomy and Data Processing, \\
Guizhou Normal University, Guiyang 550001, PR China}

\begin{abstract}
Heating of neutral gas by energetic sources is crucial for the prediction of the 21 cm signal during the epoch of reionization (EoR).
To investigate differences induced on statistics of the 21 cm signal by various source types, we use five radiative transfer simulations which have the same stellar UV emission model and varying combinations of more energetic sources, such as X-ray binaries (XRBs), accreting nuclear black holes (BHs) and hot interstellar medium emission (ISM).
We find that the efficient heating from the ISM increases the average global 21~cm signal, while reducing its fluctuations and thus power spectrum. A clear impact is also observed in the bispectrum in terms of scale and timing of the transition between a positive and a negative value. The impact of XRBs is similar to that of the ISM, although it is delayed in time and reduced in intensity because of the less efficient heating. Due to the paucity of nuclear BHs, the behaviour of the 21~cm statistics in their presence is very similar to that of a case when only stars are considered, with the exception of the latest stages of reionization, when the effect of BHs is clearly visible.
We find that differences between the source scenarios investigated here are larger than the instrumental noise of SKA1-low at $z \gtrsim 7-8$, suggesting that in the future it might be possible to constrain the spectral energy distribution of the sources contributing to the reionization process.

\end{abstract}


\section{Introduction}
\label{sec:intro}

The period during which the Universe transformed from being highly neutral to fully ionized is referred to as epoch of reionization (EoR).
Many observations, among which the absorption spectra of high redshift quasars \cite[QSOs; e.g.][]{Fan2003,Fan2006}, the cosmic microwave background radiation \cite[CMB; e.g. Planck result ][]{Planck2018} and the luminosity function of Lyman-$\alpha$ emitters \citep{Choudhury2015,Ota2017,Weinberger2019}, indicate that the Universe has been reionized at $z>5$.
However, much is still unknown about the EoR, e.g. what sources drive reionization, which are the distribution and sizes of the ionized bubbles at different stages of the EoR, and how efficiently is the neutral hydrogen heated during this period.
The 21 cm line from neutral hydrogen is considered the most promising probe of the EoR in the near future \citep{Furlanetto2006, Koopmans2015}, and it is one of the key science goals of many large radio telescopes, e.g. the Low-Frequency Array (LOFAR)\footnote{http://www.lofar.org/}, the Hydrogen Epoch of Reionization Array (HERA)\footnote{https://reionization.org/}, the Murchison Widefield Array (MWA)\footnote{http://www.mwatelescope.org/}, and the Square Kilometre Array (SKA)\footnote{https://www.skatelescope.org/}.

Different statistics of the 21 cm line have been proposed to study the EoR.
The simplest one is to measure the evolution of the global mean signal as a function of redshift. In this respect, the Experiment to Detect the Global EoR Signature (EDGES)\footnote{https://www.haystack.mit.edu/astronomy/astronomy-projects/edges-experiment-to-detect-the-global-eor-signature/} telescope has reported an absorption profile centred at 78 MHz \citep{Bowman2018}, corresponding to $z\sim 17$, which has stimulated an exciting theoretical debate on its possible origin \cite[e.g. ][]{Hektor2018, Kovetz2018, Bhatt2020}, as well as skepticism about the actual detection \citep{Hills2018}.
The power spectrum (PS), variance and skewness of the 21 cm signal are conceptually simple and can be used to constrain reionization \citep[e.g.][]{Geil2009,Harker2010,Patil2014,Kubota2016,Majumdar2016,Seiler2018,Ross2019}, as well as heating models \citep[e.g.][]{Madau1997, Christian2013, Mesinger2013, Ross2019}.
Although the 21 cm power spectrum has not been detected until now, the MWA team has published its tightest upper limit of $\Delta_{\rm 21cm} \le 1.8 \times 10^{3}\,\rm  mK^{2}$ at $k = 0.14\, h\,\rm Mpc^{-1}$ and $z = 6.5$ \citep{Trott2020}, while the LOFAR telescope has a best $2\sigma$ upper limit of $\Delta_{\rm 21cm} < 73^{2}\si{\mK}^{2}$ at $k = 0.075\, h \rm \, Mpc^{-1}$ and $z \sim 9.1$ \citep{Mertens2020}.
These measurements can already rule out some, albeit extreme, EoR models \citep{Ghara2020,Greig2020,Mondal2020}.
Higher-order statistics, such as the bispectrum \citep{Shimabukuro2016, Shimabukuro2017, Majumdar2018, Hutter2020}, the three-point correlation function \citep{Hoffmann2019}, position-dependent power spectra \citep{Giri2019}, and multi-correlations (e.g. kSZ-kSZ-21 cm correlations as in \citealt{Ma2018a} and \citealt{Plante2020}, or [CII]-[CII]-21 cm correlations as in \citealt{Beane2018}), have also been proposed to study the EoR.
As the 21 cm signal is strongly polluted by the foreground noise, a cross-correlation with other observations, such as the kinetic Sunyaev-Zel'dovich effect \citep{Jelic2010, Alvarez2015}, galaxies \citep{Lidz2009, Vrbanec2016, Moriwaki2019}, the X-ray background \citep{Liang2016, Ma2018b} and cosmic opacity \citep{Meerburg2013, Roy2020}, can reduce the statistical noise and increase the chance of a detection, in addition to offer an independent confirmation of the origin of the signal.
Besides, once tomography of the 21~cm signal will be available with the next generation of radio telescopes (see e.g. \citealt{Tozzi2000,Ciardi2003,Morales2010}), it has been suggested that modern techniques like machine learning can be used to set constraints on reionization models and reduce the degeneracies intrinsic in e.g. analysis of the power spectra \citep{Gillet2018, Mangena2020, Hassan2020}.

The amplitude and sign of the 21 cm signal crucially depends on the thermal state of the neutral gas, and, more specifically, on the value of the spin temperature in relation to the CMB temperature \citep{Field1959}.
The intergalactic medium (IGM) is expected to be cold before the EoR (i.e. during the dark ages; \citealt{Furlanetto2006}), while it becomes heated after the formation of the first sources of radiation \citep{Pritchard2007, Mesinger2013, Pacucci2014}.
While stars drive the reionization process, the heating (and partial ionization) of the gas responsible for the 21 cm signal is determined by more energetic sources (see e.g. \citealt{Eide2018,Eide2020}).
X-ray binaries (XRBs) were thought to be the main heating source \citep[e.g.][]{Mesinger2011}, but the efficiency of such heating is still debated (see e.g. \citealt{Fragos2013a, Fragos2013b,Fialkov2014}).
Other sources, such as accretion on nuclear black holes in galaxies (e.g. QSOs) and emission from the shock heated interstellar medium (ISM) can also heat the neutral IGM \citep{Eide2018, Eide2020}.
{However, the contribution of these sources to heating and ionization is still debated \cite[e.g. ][]{Pacucci2014, Cohen2018}, introducing additional degeneracies of parameters in comparison to models with stars only.}

In this paper, we adopt the simulations from \cite{Eide2018, Eide2020}, which have modeled the properties of energetic sources (such as XRBs, nuclear black holes and hot ISM) on the basis of cosmological hydrodynamical simulations and multi-frequency 3D Monte Carlo radiative transfer, to study how different 21 cm statistics (global mean, variance, skewness, power spectrum and bispectrum) are affected by such sources.
{Although \cite{Eide2018, Eide2020} made a comparison to some available observations, no estimate of the 21~cm signal was made.}
This paper is structured as follows.
In Section 2, we describe the simulations and post-processing techniques adopted in this work.
Section 3 presents the method of computing the 21 cm signal.
The results of the 21 cm statistics and their detectability are shown in Section 4, while Section 5 contains the discussion and conclusions.

\section{Simulations of Cosmic Reionization}

Here we employ the simulations of reionization obtained by post-processing the high resolution cosmological hydrodynamical simulation Massive Black-II \citep[MBII; ][]{Khandai2015} with the 3D radiative transfer (RT) code CRASH \citep{Ciardi2001,Maselli2009,Graziani2013,Graziani2018,Glatzle2019}.
The simulations have been extensively discussed in \citet{Eide2018} and \citet{Eide2020}, hereafter E18 and E20 respectively, and have also been used to study e.g. the cross-correlation between the 21 cm signal and the X-ray background \citep{Ma2018b}, or the [OIII]
emitters \citep{Moriwaki2019}, as well as the modelling and observability of the $^3$He$^+$ line from singly ionized helium \citep{Khullar2020}.
For more details about these simulations we refer the reader to \cite{Khandai2015}, E18 and E20, while here we just briefly summarize their main characteristics.

The MBII simulation has been run in a box of length $100\,h^{-1} \,\rm cMpc$ with $2\times 1792^{3}$ gas and dark matter particles, corresponding to a resolution of $2.2 \times 10^6 h^{-1}$~M$_\odot$ and $1.1 \times 10^7 h^{-1}$~M$_\odot$, respectively.
The cosmological parameters adopted are  $\Omega_{\Lambda}= 0.725$, $\Omega_{m} = 0.275$, $\Omega_{b} = 0.046$, $h = 0.701$, $n_{s} = 0.968$ and $\sigma_{8} = 0.816$ \citep[Wilkinson Microwave Anisotropy Probe 7 result,][]{Komatsu2011}.
The simulation follows the evolution and properties of stellar populations, galaxies and black holes (BHs), such as mass, age, metallicity, star formation rate and accretion rate of BHs.
{Star formation is computed through a subgrid model based on a multiphase description of star-forming gas, so that only cold gas forms stars.
Star particles are probabilistically created from gas particles according to their star formation rates.
Haloes are identified with a friends-of-friends procedure (resulting in a minimum halo mass of $\sim 5 \times 10^8 h^{-1}$~M$_\odot$), while galaxies are defined with at least 64 star particles.

Based on the stellar, galactic and BH properties}, E18 and E20 modeled the luminosity and spectrum of ionizing and heating sources, such as stars \citep{Eldridge2012}, XRBs \citep{Fragos2013b, Madau2017}, supernova heated ISM \citep{Mineo2012_ism, Pacucci2014}, and accreting nuclear BHs hosted in galaxies \citep{Shakura1973, Krawczyk2013}.
{Briefly, the spectra of star particles are modeled from the 2012 version of the population synthesis code BPASS \citep{Eldridge2012} depending on their stellar mass, age and metallicity of a single star population.
The luminosity of a BH is scaled by its accretion rate with an efficiency parameter $0.1$, while
the spectrum adopted for all BHs is obtained from averaging observations of 108,104 low-$z$ QSOs \citep{Krawczyk2013}, and it is a broken power law with a spectral index of 1 for photons of energies above $ 200\,\rm eV$.
The luminosities and spectra of XRBs are scaled with galactic physical properties according to the libraries of \cite{Fragos2013b} and \cite{Madau2017}.
The luminosity of the heated ISM is scaled with the star formation rate by following \cite{Mineo2012_ism}, while the spectrum is flat until thermal energy break at 240 eV and a broken power-law with a spectral index of 3 at higher energies.
We note that all the values adopted to model the sources are either taken directly from the MBII simulations or from the literature, without the adoption of further parameters.}
{\cite{Khandai2015} and E20 have compared the properties of galaxies from the MBII simulation to observations, finding consistency with respect to e.g. the shape of the cosmic spectral energy distribution of galaxies, the galaxy stellar mass function, the quasar bolometric luminosity function, and the luminosity function of galaxies at AB magnitude.}

The physical properties of the IGM, as well as the location and properties of the sources, as derived from MBII are mapped onto a 256$^3$ Cartesian grid and used by CRASH to evaluate the redshift evolution of the ionization and temperature state of the IGM from $z=20$ to $z=5$ under different combinations of these sources.
Note that the effect of the different source spectra is accurately calculated by the multi-frequency RT as they are sub-divided into 82 frequency bins between 13.6~eV and 2~keV.
Here we analyze the results of five simulations with a constant escape fraction of UV photons $f_{esc}=0.15$, including: only the stars in galaxies (GAL), stars and ISM (GALISM),  stars and accreting nuclear BHs (GALQSO), stars and XRBs (GALXRB), and all sources combined (GXQI).

\begin{figure}
    \centering
    \includegraphics[width=0.95\linewidth]{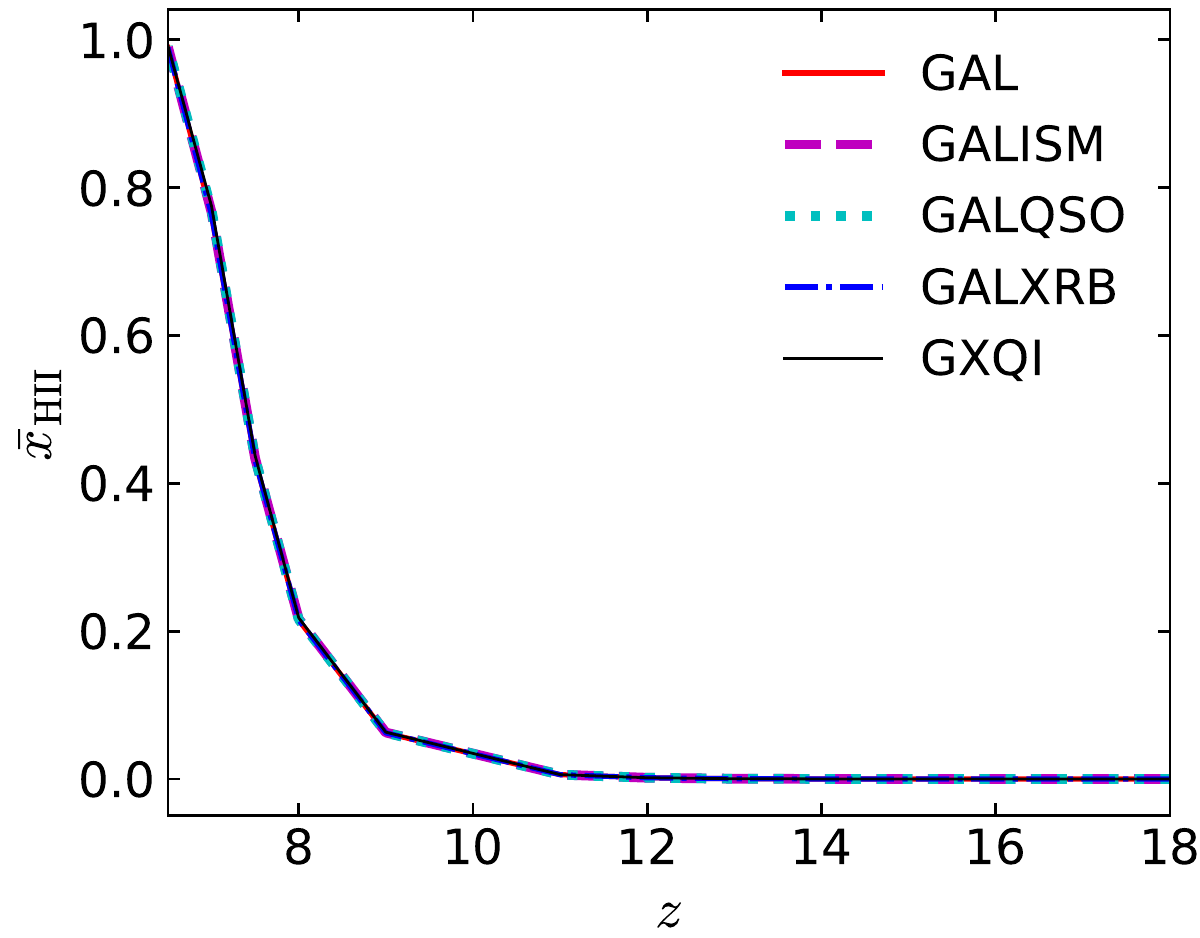}
    \caption{Redshift evolution of the volume averaged HII  fraction in model GAL (solid red line), GALISM (dashed magenta), GALQSO (dotted cyan), GALXRB (dash-dotted blue) and GXQI (solid black).
    }
    \label{fig:xhii_history}
\end{figure}
As a reference, in Fig.~\ref{fig:xhii_history} we show the evolution of the volume averaged ionization fraction, $\bar{x}_{\rm HII}$, for the five simulations.
E18 and E20 found that, as expected, the reionization history is dominated by stars, while more energetic sources only produce partially ionized, warm gas.
For this reason, the curves in the figure are basically the same in all models, with $\bar{x}_{\rm HII} = 0.5$ at $z\approx 7.5$ and $\bar{x}_{\rm HII} = 0.99$ at $z\approx 6.5$.
However, energetic sources quickly heat up the gas, with the five models showing clear differences in the temperature distributions (see E18 and E20 for more details).
Thus, also the 21 cm signal is expected to vary among the models.

Here we note that, as numerical simulations on large scales are not able to resolve the ionization front for short mean free path photons, such as in the case of the pure UV spectrum in the GAL scenario \citep{Ross2017}, the cells that contain the front can falsely appear as partially ionized and warm (see also discussion in E18 and E20), leading to wrong predictions of the 21 cm signal.
To correct for this effect, we adopted the post-processing technique described in \cite{Ma2020}.
More specifically, each partially ionized cell is divided into 8 sub-cells that are either fully ionized or completely neutral, depending on their minimal distance from the fully ionized cells calculated with the Euclidean distance transform  \citep{Rosenfeld1966,Busch2020}.
The temperature of the neutral sub-cells is set to that of its closest neutral cell which is not in direct contact with fully ionized cells, while the temperature of ionized cells is the same one of the parent cell (the exact value does not impact the 21 cm signal in this case).
{ Appendix B shows how the correction affects the 21 cm power spectra and bispectra.}

\section{21 cm Signal}

The differential brightness temperature (DBT) of the 21 cm signal, $\tto$, at redshift $z$ can be  computed as \citep{Furlanetto2006}:
\begin{equation}
\label{eq:21cm_original}
    \tto (z) = \frac{T_{s}-T_{\rm CMB}}{1+z}(1-e^{-\tau}),
\end{equation}
where $T_{s}$ is the spin temperature of neutral hydrogen, $T_{\rm CMB}$ is the CMB temperature, and $\tau$ is the integrated opacity of the 21 cm signal through the intervening gas:
\begin{equation}
    \tau = \frac{(1+z)\Psi (z)  (1+\delta)x_{\rm HI}}{T_{s} \phi(z)},
\end{equation}
where $\delta$ is the gas over-density, $x_{\rm HI}$ is the neutral fraction of hydrogen, the coefficient factor
\begin{equation}
    \Psi (z) = \SI{27}{\milli\K}\, \frac{\Omega_{b}h^{2}}{0.023} \sqrt{\frac{0.15}{\Omega_{m}h^{2}}  \frac{1+z}{10}},
\end{equation}
and the velocity term $\phi(z) = 1+\frac{{\rm d}v_{\|}/{\rm d}r_{\|}}{H_{z}}$, with $H_{z}$ Hubble expansion rate at $z$, and ${\rm d}v_{\|}/{\rm d}r_{\|}$ gradient of the peculiar velocity along the line of sight.
As we mainly focus on the heating and ionization phase of the EoR, we assume that the Ly$\alpha$ photons already couple the spin temperature to the kinetic temperature, i.e. $T_{s} = T_{k}$.
When $\tau \ll 1$ (i.e. optical thin), Eq.~\ref{eq:21cm_original} can be simplified as:
\begin{equation}
\label{eq:21cm_thin}
    \tto (z) =  \frac{\Psi (z) (1+\delta)x_{\rm HI}}{\phi(z)}  \left(1-\frac{T_{\rm CMB}}{T_{s}}\right).
\end{equation}

The 21 cm power spectra \citep{Mao2012,Majumdar2013, Ross2020} and bispectra \citep{Majumdar2020} can be significantly affected by redshift space distortions (RSD), in particular during the early stages of the EoR.
For example, the 21 cm power spectra can be increased by a factor larger than 4 at $\bar{x}_{\rm HII} \sim 0.2$ and $k \sim 0.2\, h\,\rm Mpc^{-1}$, while the effect is negligible towards the end of the EoR \citep{Mao2012}.
The effect on the bispectrum is even larger, as RSD can change both its sign and its magnitude.
\cite{Majumdar2020} find that, in the redshift range covered by their study, i.e. $z=7-13$, the magnitude is increased/decreased by $50-100\%$ for small and intermediate $k$-triangles without sign changing, while by $100-200\%$ with sign change for larger $k$-triangles.
Thus, this effect should be properly modelled for a correct evaluation of the 21 cm signal throughout the full EoR.
Here we use the MM-RRM scheme (Mesh-to-Mesh Real-to-Redshift-Space-Mapping) described in \cite{Mao2012} to correct for the RSD caused by peculiar velocities.
The MM-RRM scheme redistributes the 21 cm DBT from the real space to the redshift space by shifting the cell boundaries to their locations in redshift space according to the velocity along the line of sight.
Thus the real-space cells are stretched or compressed in redshift space depending on the velocity.
\cite{Mao2012} found that the term $\phi(z)$ is cancelled in the case of optical thin gas (eq.~\ref{eq:21cm_thin}), i.e. the velocity shift is much more important than the $\phi(z)$ term when computing the 21 DBT in redshift space.

\section{Results}
\label{sec:res}

\begin{figure*}
    \centering
    \includegraphics[width=0.95\linewidth]{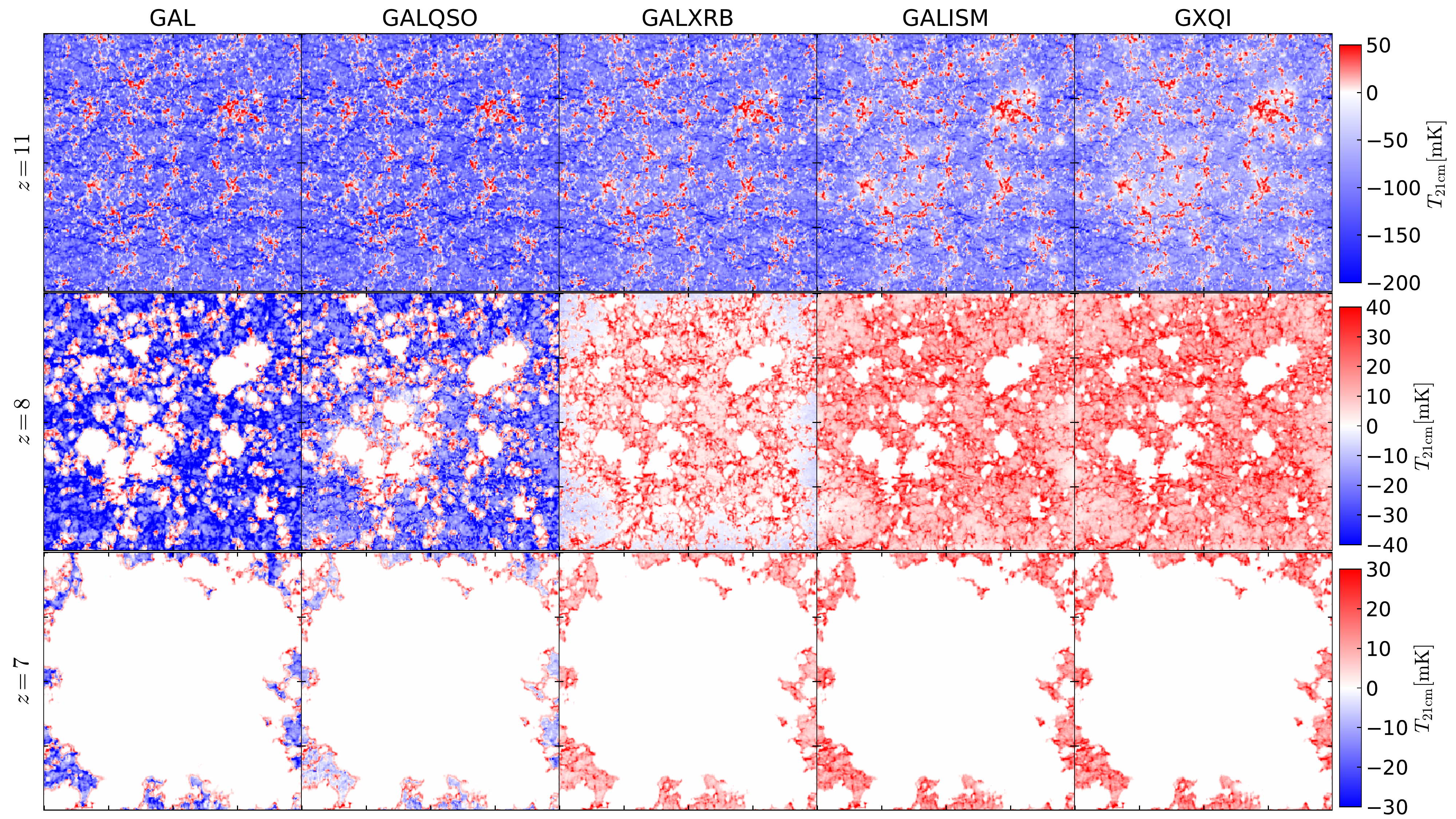}
    \caption{Slices of thickness $0.4h^{-1}\mathrm{cMpc}$ and side length 100 $h^{-1}$ cMpc showing the 21 cm differential brightness temperature in models GAL, GALQSO, GALXRB, GALISM and GXQI (from left to right), at $z = 11$ (top, $\bar{x}_{\rm HII} = 0.01$), $z = 8$ (middle, $\bar{x}_{\rm HII} = 0.21$) and $z = 7$ (bottom, $\bar{x}_{\rm HII} = 0.76$).}
    \label{fig:s_21cm_image}
\end{figure*}
While we refer the reader to E18 and E20 for a better understanding of the differences caused by the more energetic sources, here we emphasize that stars dominate the emissivity at frequencies $\nu <\SI{60}{\eV}$ (UV photons), while the hot ISM becomes the main emitter in the range $[60-500]\,\si{\eV}$ (hard UV and soft X-ray photons) and the XRBs dominate at $\nu >500 \, \rm eV$ (X-ray photons). Finally, the nuclear BHs, whenever present, have the strongest emission at all frequencies, but, due to their paucity, they have a negligible effect on the global ionization budget.
The spectra of all the sources, as well as their contribution to the total emissivity, can be found in E18 and E20.

As a visual reference, in Fig.~\ref{fig:s_21cm_image} we show maps of the 21 cm DBT in our five models.
As mentioned earlier, full H reionization is driven by stellar type sources and thus the ionized regions (corresponding to $\tto = \SI{0}{mK}$) are basically the same in all models, with a slightly larger extent observed in the presence of nuclear BHs. The amplitude and distribution of the non-zero 21 cm DBT, though, are clearly different.
At $z=11$, the red parts (corresponding to the signal in emission) are similar in all models and are due to the partially ionized and warm gas found in the hydrodynamic simulations, i.e. at this time the effect of the ionizing radiation on the signal in emission is still negligible. On the other hand, we can clearly see differences, albeit small, in the signal in absorption (corresponding to the blue parts). These are due to the presence of energetic sources, which, having different spectral energy distributions, are more or less efficient at heating the gas.
As the hot ISM is very efficient at uniformly heating the IGM (see E18 and E20), at $z=8$ both the GALISM and GXQI models have a signal which is fully in emission. Conversely, XRBs have not yet heated all the IGM up to temperatures $T_{k} > T_{\rm CMB}$, and thus some cells in the GALXRB model are still in absorption, while those in emission have an amplitude of the signal lower than in the GALISM and GXQI models.
Most IGM in the GAL and GALQSO models is still cold, although heating from BHs starts to be non-negligible (see also \citealt{Baek2010,Ross2019}). The signal visible in emission is mainly due to the effect of the few hard-UV photons emitted by the stellar sources. The same signal can be also seen at $z=7$.
While most of the IGM at $z=7$ is fully ionized, some partially ionized warm cells are still present which produce a signal in emission in the GALISM, GALXRB and GXQI models, while the same cells are still cold in the GAL and GALQSO models.

In the following, we will study the impact of energetic sources on the 21 cm global signal, power spectrum and bispectrum.

\subsection{0-D and 1-D statistics of 21 cm DBT}
\begin{figure*}
    \centering
    \subfigure{
    \begin{minipage}{0.45\linewidth}
    \centering
    \includegraphics[width=1\linewidth]{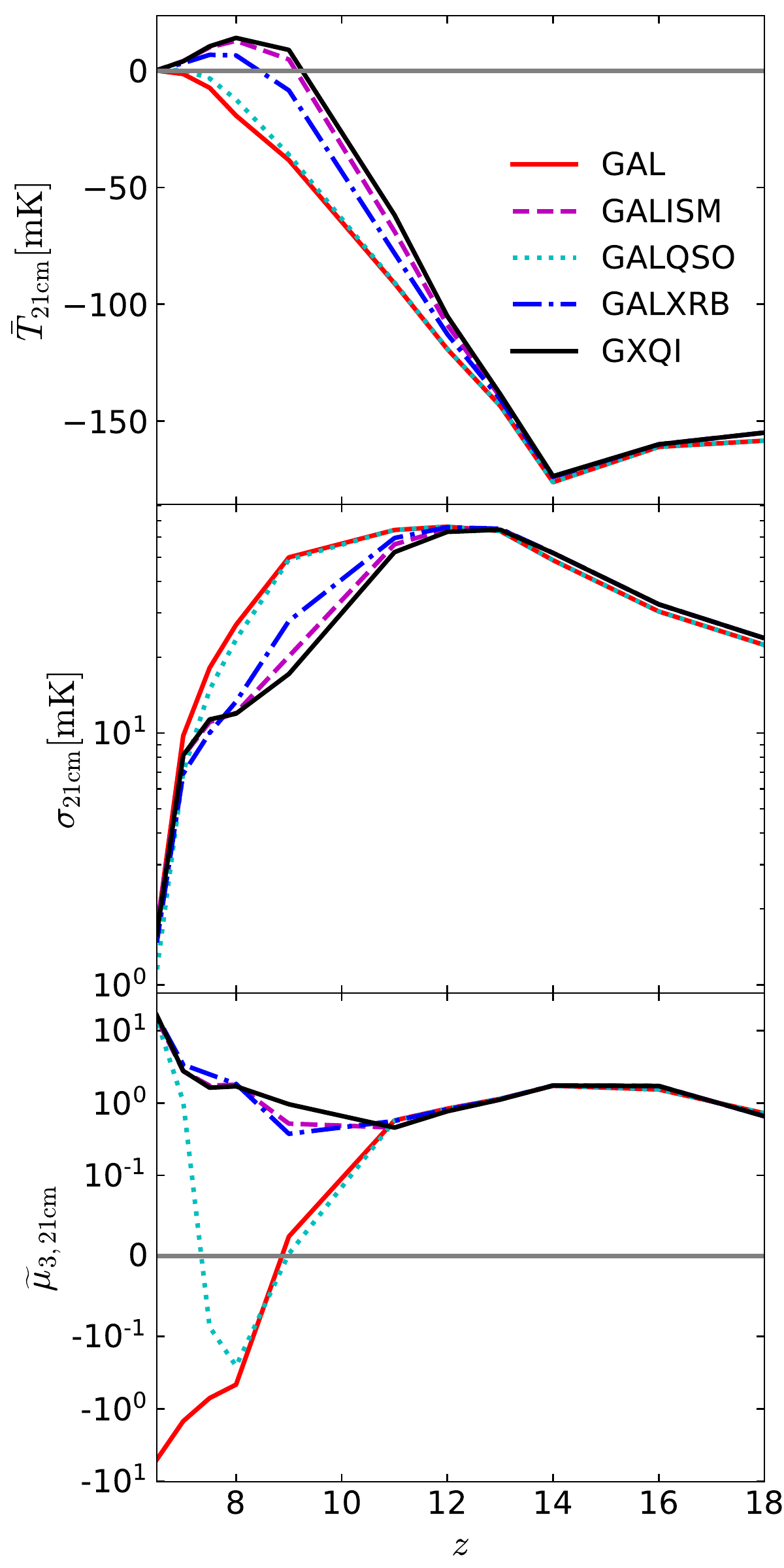}
    \end{minipage}
    }
    \subfigure{
    \begin{minipage}{0.45\linewidth}
    \centering
    \includegraphics[width=1\linewidth]{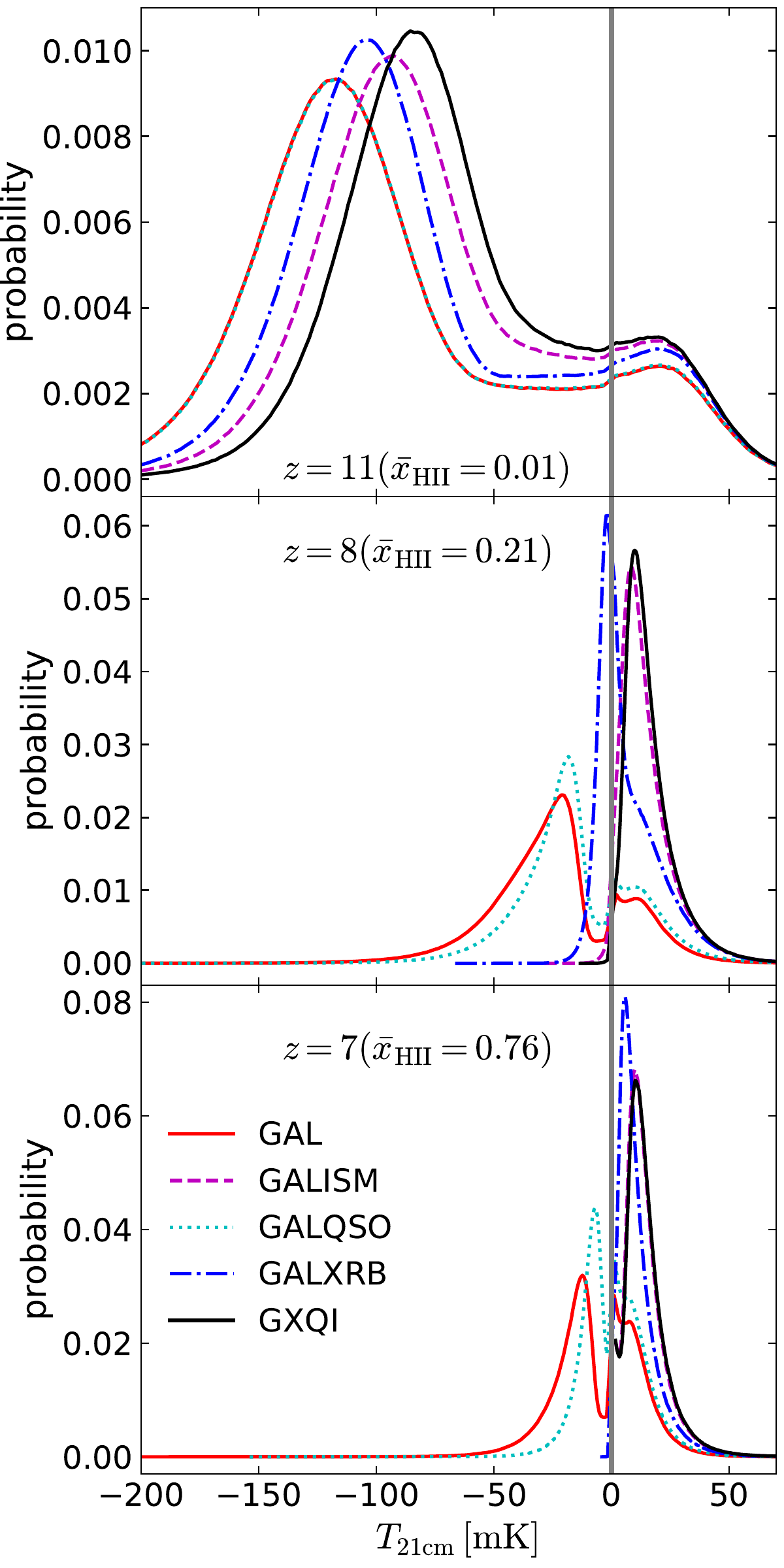}
    \end{minipage}
    }
    \caption{{\bf Left:} Redshift evolution of volume averaged $\tto$ (\meantto, top panel), 21 cm standard deviation (\stdto, central panel) and  skewness (\skewtto, bottom panel) in model GAL (solid red), GALISM (dashed magenta line), GALQSO (dotted cyan), GALXRB (dash-dotted blue) and GXQI (solid black).
    The horizontal gray lines in the top and bottom panels are drawn at zero to guide the eye.
    {\bf Right:} From top to bottom, probability density distributions of $\tto$ in five models having the same line types with the left at $z = 11$ ($\bar{x}_{\rm HII} = 0.01$), 8 ($\bar{x}_{\rm HII} = 0.21$) and 7 ($\bar{x}_{\rm HII} = 0.76$).
    The vertical gray lines are drawn at zero to guide the eye.
    Note that the cells with $\tto \sim 0$~mK are removed.}
    \label{fig:s21cm_1d_ts}
\end{figure*}

The left panel of Fig.~\ref{fig:s21cm_1d_ts} shows the volume averaged DBT, $\meantto$, the standard deviation, $\stdto$, and the skewness of the 21 cm signal in the five models.
These quantities might be measured in the near future and used to study the physics of the EoR (see e.g. \citealt{Kubota2016}). The skewness indicates how a distribution is skewed around the mean value: if it is larger (smaller) than zero, it means that there is more weight in the right (left) tail of the distribution. Here, the skewness is defined as
\begin{equation}
    \skewtto = \frac{\mu_3\left(\tto\right)}{\mu_2\left(\tto\right)^{3/2}} = \frac{(\tto - \meantto)^{3}}{\left(\stdto\right)^3},
\end{equation}
with the $i$-th central moment $\mu_i$.

Although the energetic sources (except nuclear BHs) are present from the beginning of the simulations (at $z=20$),  all models have similar mean, variance and skewness at $z \gtrsim 13$, due to their very low impact in the early stages of the EoR.

Without substantial heating of the IGM, the gas temperature $T_{k} \ll T_{\rm CMB}$, and thus all models have $\meantto \ll 0$~mK at the beginning of reionization, e.g. it is about -170~mK at $z=14$.
At $z>14$, $\meantto$ decreases with decreasing $z$ due to the adiabatic expansion of the Universe, which dominates the temperature evolution until shock heating first and then radiative heating become relevant.
As stars emit a negligible amount of high-energy photons, the GAL model has $\meantto < 0$~mK throughout the EoR, becoming close to zero towards its end.
As more energetic sources are able to heat the neutral gas, models GALISM, GALXRB and GXQI show much higher $\meantto$ at $z \lesssim 13$.
Once the neutral hydrogen is heated at temperatures $T_{k} > T_{\rm CMB}$, the 21 cm signal becomes in emission, i.e. $\meantto > 0$~mK. The timing of such transition is sensitive to the source model.
As accreting nuclear BHs appear only at $z<13$ in the simulations and they are rare at redshifts relevant to the EoR, the GALQSO model has a $\meantto$ which is just slightly higher than the one of the GAL model, and it becomes positive only at $z \lesssim 7$.
Both the XRBs and the hot ISM are instead more ubiquitous and thus more effective at heating the IGM. The XRBs emit predominantly X-rays, which are less efficient than the hard UV and soft X-ray photons emitted by the ISM. As a consequence, the transition to the signal in emission happens at $z \lesssim 8$ and at $z \lesssim 9$ for the GALXRB and GALISM models, respectively. As the GXQI model includes all sources, it displays the highest $\meantto$.

At $z \gtrsim 13$, when the heating is negligible, the fluctuations of the 21 cm signal are similar in the five models and are dominated by the over-density and hydro-temperature, thus $\stdto$ increases with decreasing redshift.
Because of the long mean free path of high-energy photons, in addition to increasing the gas temperature and the DBT, energetic sources render the heating more uniform, hence reducing the fluctuations (see also the power spectra in  Fig.~\ref{fig:s_21cm_ps_vk}) and the amplitude of $\stdto$ at $z \lesssim 13$.
As a consequence, the five simulations present different values of $\stdto$, which is highest in the GAL model and lowest in the GXQI one.
As the spectrum of the XRBs is much harder than that of the hot ISM,  the $\stdto$ of the GALXRB model is lower than that of the others at $z \sim 7.5$.

In all models, at $z>11$ most of the IGM is cold, with a $\meantto \lesssim -100$~mK. For this reason, the several cells heated above the CMB temperature by hydrodynamic processes (see the right panel of Fig.~\ref{fig:s21cm_1d_ts}) bias the DBT distribution towards the high temperature tail, resulting in a positive skewness, similar for
all models.
At $z<11$, the heating from the energetic sources still induces a bias towards the high temperature tail of the 21~cm DBT distribution, resulting in a positive \skewtto in the GALISM, GALXRB and GXQI models.
In the absence of such sources, though, as $\meantto$ steadily increases with decreasing redshift, a bias appears instead towards the low temperature tail, so that the \skewtto of the GAL model decreases quickly and becomes negative at $z<9$.
The nuclear BHs have a significant contribution only towards the end of the reionization process, so that the \skewtto of the GALQSO model is similar to that of the GAL model at $z>8$, but it quickly increases at $z<8$ and approaches the values of the GALISM, GALXRB and GXQI models at $z<7$.

The right panel of Fig.~\ref{fig:s21cm_1d_ts} presents the volume weighted probability density distribution functions (PDF) of the 21 cm signal in the five models at $z=11$, 8 and 7.
For the ionization fraction and temperature distributions, we refer the reader to E18 and E20.
To make the plot more readable, we are not showing the fraction of cells corresponding to $\tto=0$, which approximately equals the volume averaged ionization fraction, i.e. 0.01 at $z=11$, 0.21 at $z=8$ and 0.76 at $z=7$.
At $z=11$, when the heating of X-ray sources is still very weak but non-negligible, the 21 cm PDF shows two peaks in all models.
One is dominated by the temperature as determined by the hydrodynamic simulation, and thus its location is the same in all models (i.e. at $\tto \sim 20$~mK).
The other peak, instead, has a negative value, as at $z=11$ the heating of the predominantly neutral IGM is still very limited. The effect of the energetic photons, though, is nevertheless visible, as the distributions peak at $\tto \sim -120$~mK in the GAL and GALQSO models, while the GALXRB, GALISM and GXQI models peak at $\tto \sim -105$~mK, $\tto \sim -95$~mK and $\tto \sim -85$~mK, respectively.
These common features disappear when the heating of the predominantly neutral IGM becomes strong enough, e.g. at $z=8$ and 7 in the GALXRB, GALISM and GXQI models.
Because of the harder spectra of XRB sources, the $\tto$ of the peak in the GALXRB model is lower than that of GALISM and GXQI models at $z=8$ and 7, and it is still negative at $z=8$, while it gets closer to the latter cases at $z=7$.
Once the IGM gas is heated at temperatures $T_{k}\gg T_{\rm CMB}$, e.g. in the GALISM and GXQI models at $z=7$, the 21 cm PDFs become the same as they are now independent from the gas temperature.
As in the GAL model the spectrum of the stellar type sources  has a weak hard-UV tail up to $\sim \SI{100}{\eV}$ (see Fig.~1 in E20), this can heat up some gas around the fully ionized regions (see also Fig.~\ref{fig:s_21cm_image}), inducing the small emission peaks observed in the GAL and GALQSO models at $z=8$ and 7.
While at the highest redshifts the influence of the BHs is not visible, their effect in terms of heating and ionization becomes obvious below $z \sim 8$, with the emission peak being larger than in the GAL simulations and the absorption peak being shifted towards larger values.

We note that the $\tto$ deviation, skewness and 1-D PDF shown here are the statistics of cells from the simulations, while the measured ones would depend on the configuration of telescope arrays e.g. the angular resolution. We will discuss the observability of the signal later on.

\subsection{21 cm Power Spectra}

The power spectrum of the 21 cm signal is defined as:
\begin{equation}
P_{\rm 21cm}(\bm{k}) = \delta_{D}(\bm{k}+\bm{k'}) \langle \tto(\bm{k})\tto(\bm{k'})\rangle
\end{equation}
where $\delta_{D}$ is the Dirac function, $\tto(\bm{k})$ is the 21 cm DBT in the Fourier space, and the angle bracket means the ensemble average.
In the following, we will present results in terms of the normalized form $\Delta_{\rm 21cm} = k^{3}/2\pi^{2} \times P_{\rm 21cm}$.
Note that, limited by the scale of the simulation box, the  results of the 21 cm power spectra and bispectra in the next subsection are not robust at $k<0.3\,\rm Mpc^{-1}$, as they become sample variance dominated.

\begin{figure*}[th]
    \centering
    \subfigure{
    \begin{minipage}{0.45\linewidth}
    \centering
    \includegraphics[width=1.0\linewidth]{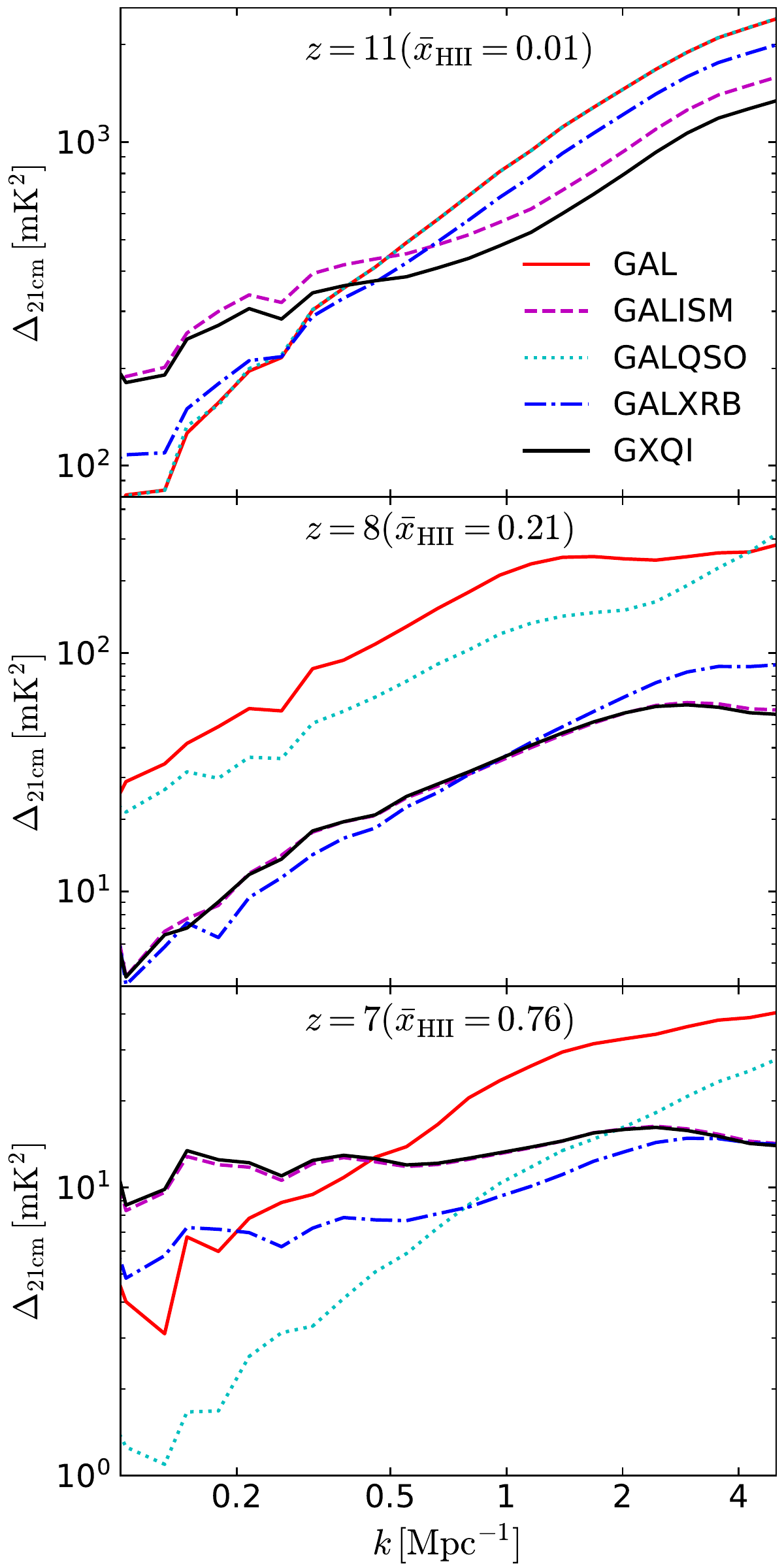}
    \end{minipage}
    }
    \subfigure{
    \begin{minipage}{0.45\linewidth}
    \centering
    \includegraphics[width=1.0\linewidth]{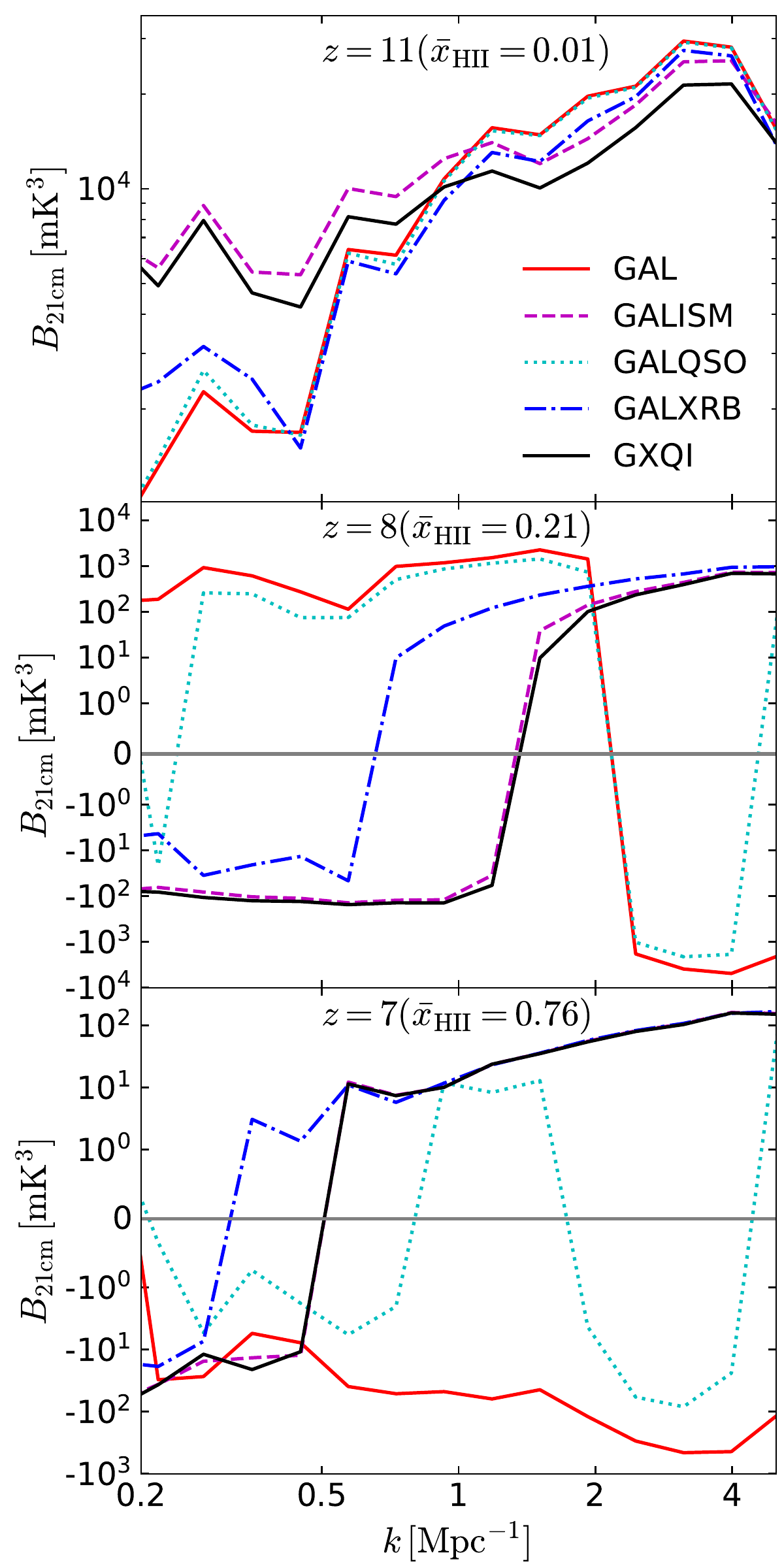}
    \end{minipage}
    }
    \caption{Power spectra (left panel) and bispectra of equilateral triangles (right panel) of the 21 cm signal in model GAL (solid red lines), GALISM (dashed magenta), GALQSO (dotted cyan), GALXRB (dash-dotted blue) and GXQI (solid black).
    From top to bottom, the panels refer to $z = 11$ ($\bar{x}_{\rm HII} = 0.01$), $z = 8$ ($\bar{x}_{\rm HII} = 0.21$) and $z = 7$ ($\bar{x}_{\rm HII} = 0.76$).
     The horizontal gray lines are drawn at zero to guide the eye.
}
    \label{fig:s_21cm_ps_vk}
\end{figure*}
The left panel of Fig.~\ref{fig:s_21cm_ps_vk} shows $\Delta_{\rm 21cm}$ at $z=11$, 8 and 7 in our five models.
During the early stages of the EoR (at $z \gtrsim 13$), the fluctuations of $\tto$ are mainly dominated by the over-density of gas matter and the inhomogeneous hydro-temperature. Once the impact of energetic sources becomes relevant (e.g. at $z=11$), the 21~cm power spectra in the GAL and GALQSO models is larger than in the other models on the smallest scales (i.e. at $k \gtrsim 0.5\,\rm Mpc^{-1}$) because the radiation from the energetic sources increases the gas temperature $T_{k}$ in their vicinity, reducing the fluctuations of $\tto$ (as seen in Fig.~\ref{fig:s21cm_1d_ts}), as well as the amplitude of $\Delta_{\rm 21cm}$. The reduction is largest (smallest) in the GXQI (GALXRB) model, reflecting the strength of the heating.
On larger scales, while the 21~cm signal in the GAL and GALQSO models is still dominated by the gas-overdensity because the IGM is mostly neutral and cold, in the other models the effect of the inhomogeneous heating becomes relevant (see also Fig.~\ref{fig:s_21cm_image} and \citealt{Pritchard2007}) and increases the amplitude of the power spectrum.
We note that, due to the hardness of their spectrum, the XRBs are less effective at heating than the hot ISM, and thus the GALXRB model resembles more closely the models without X-ray sources.
At $z=8$ the nuclear BHs' contribution to local ionization and heating (in terms of larger fully ionized regions and partial ionization and heating outside them) is evident, resulting in a $\Delta_{\rm 21cm}$ lower than in the GAL model, but still higher than in the other models, in which the more diffuse partial ionization and heating further reduce the amplitude of the power spectra. Because the GALISM and GXQI models have a similar IGM temperature and ionization fractions (see e.g. Fig.~9 and 10 in E20), and thus a similar 21~cm signal (see Fig.~\ref{fig:s_21cm_image}), they also have the same $\Delta_{\rm 21cm}$.
The power spectrum of the GALXRB model is similar as well, but, because the heating from XRBs is weaker and less diffuse than that from the ISM, this results in an amplitude which is slightly lower (higher) than the one in the GALISM model at $k<1\,\rm Mpc^{-1}$ ($k>1\,\rm Mpc^{-1}$).
At $z=7$ similar considerations apply. Now, though, the effect of nuclear BHs is even stronger, resulting in a larger difference with the GAL model.
As at this stage on small scales the heating and ionization from XRBs has become similar to those of the ISM, the GALXRB, GALISM and GXQI all have the same $\Delta_{\rm 21cm}$. On large scales, though, the differences observed at higher redshift remain, with an amplitude in the GALXRB model lower than in the others because of the typically lower values of the 21~cm DBT (see the right panel of Fig.~\ref{fig:s21cm_1d_ts}).
\begin{figure*}
    \centering
    \subfigure{
    \begin{minipage}{0.45\linewidth}
    \centering
    \includegraphics[width=1.0\linewidth]{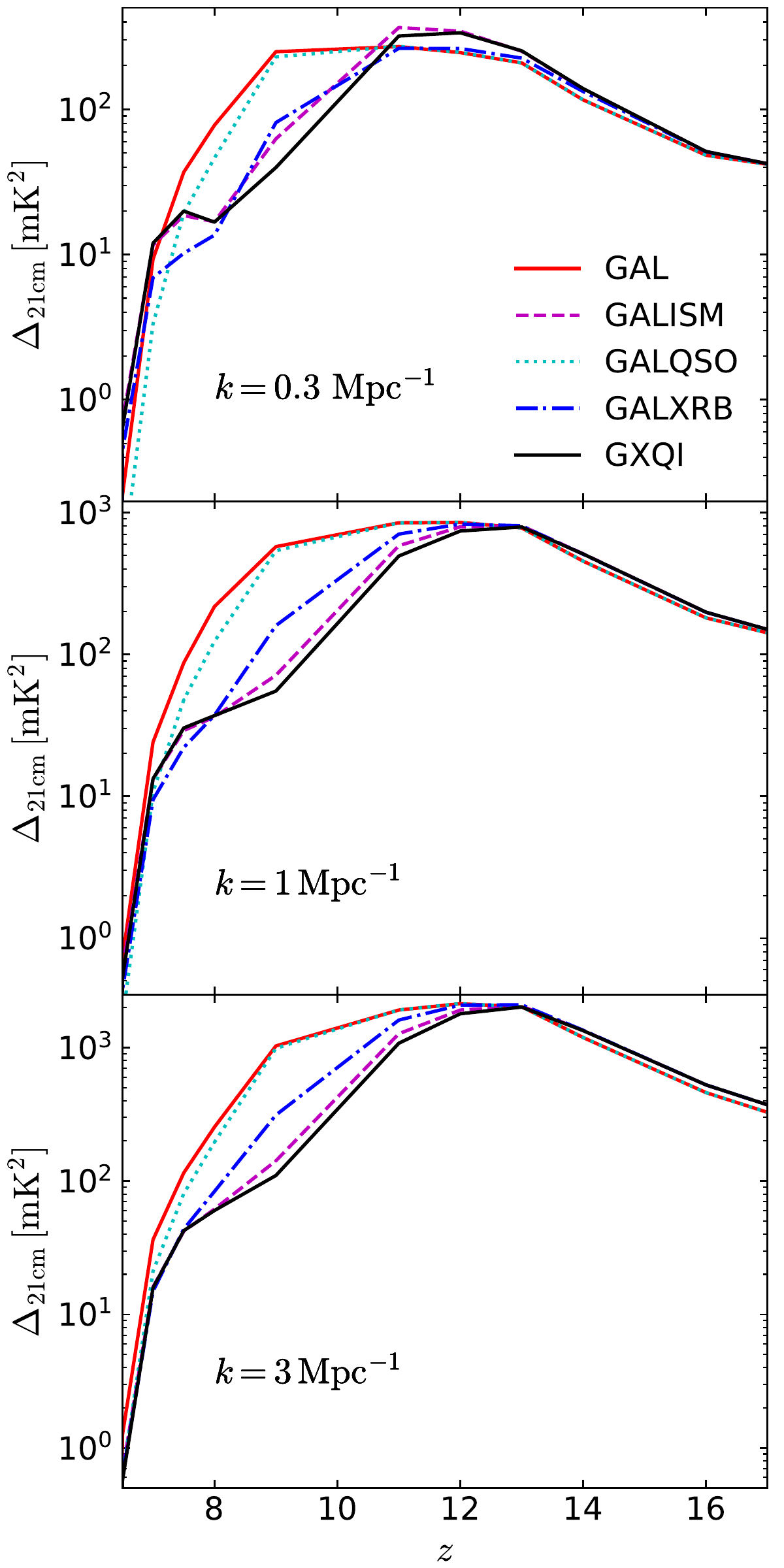}
    \end{minipage}
    }
    \subfigure{
    \begin{minipage}{0.45\linewidth}
    \centering
    \includegraphics[width=1.0\linewidth]{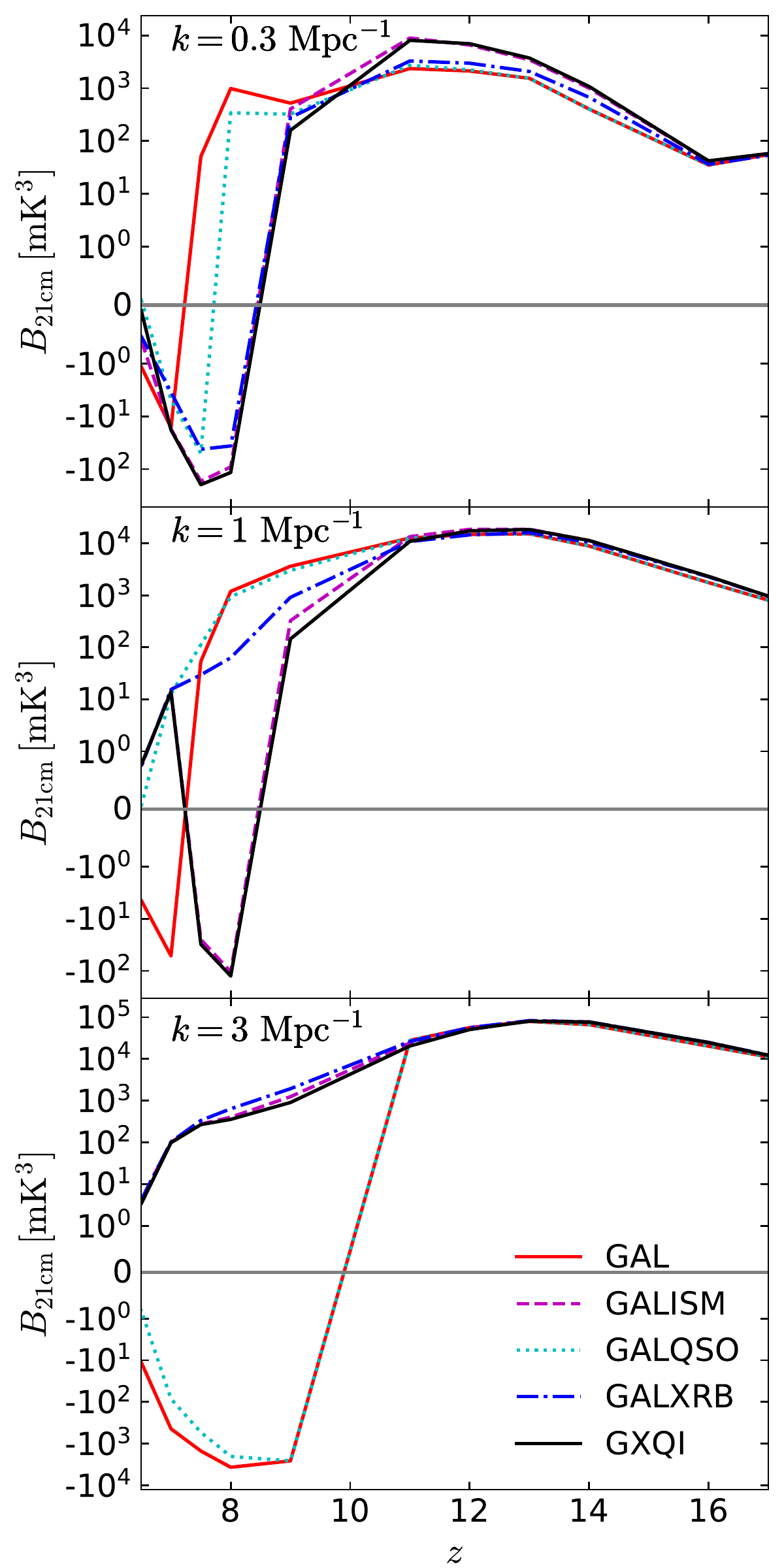}
    \end{minipage}
    }
    \caption{Redshift evolution of 21 cm power spectra (left) and bispectra of equilateral triangles (right). From top to bottom, $k = 0.3$, 1 and 3 Mpc$^{-1}$.
     In both panels, the lines refer to the model GAL (solid red), GALISM (dashed magenta), GALQSO (dotted cyan), GALXRB (dash-dotted blue) and GXQI (solid black).
     The horizontal gray lines are drawn at zero to guide the eye.
     }
    \label{fig:s_21cm_ps_vz}
\end{figure*}

The left panel of Fig.~\ref{fig:s_21cm_ps_vz} shows the evolution of $\Delta_{\rm 21cm}$ at $k = 0.3$, 1 and 3 Mpc$^{-1}$ as a function of redshift in the five models.
The profiles closely resemble those of the standard deviation of $\tto$ shown in the left panel of Fig.~\ref{fig:s21cm_1d_ts}.
At $z \gtrsim 13$, before the impact of X-ray sources on the 21~cm signal becomes significant and the signal is dominated by the gas over-density and hydro-temperature distribution, the five models show a similar $\Delta_{\rm 21cm}$ at all scales.
At $z \lesssim 13$, as reionization proceeds, $\Delta_{\rm 21cm}$ decreases with decreasing $z$, and a more powerful heating (together with more diffuse partial ionization) leads to smaller amplitudes of $\Delta_{\rm 21cm}$, with the largest differences between models observed at $z \sim 9$.
More specifically, the GXQI and GALISM models have the lowest $\Delta_{\rm 21cm}$, while the GAL and GALQSO models have the largest, e.g. by a factor of $\sim$ 10 at $z=9$ and $k =$ 3 Mpc$^{-1}$.
The $\Delta_{\rm 21cm}$ of the GALXRB model is in between them.
The GALISM and GXQI models show clear differences at $z>8$, while once the neutral and partially ionized hydrogen is heated to temperatures $T_{k}\gg T_{\rm CMB}$, their $\Delta_{\rm 21cm}$ converges to the same values.
At $k=$ 1 Mpc$^{-1}$, the profiles are similar to those at $k=$ 3 Mpc$^{-1}$, except that the $\Delta_{\rm 21cm}$ of the GXQI and GALISM models are higher than that of the GALXRB model at $z<8$.
At $k=$ 0.3 Mpc$^{-1}$, this feature is more significant, due to the growth of 21~cm fluctuations caused by the ionized bubbles, whose characteristic size directly relates to the scale of the fluctuations (see e.g. \citealt{Furlanetto2006}).
However, the higher $\Delta_{\rm 21cm}$ of the GXQI and GALISM models at $z>11$ is due to the effect of non-uniform heating at high-$z$.
Besides, the GALXRB and GALISM models display similar $\Delta_{\rm 21cm}$ at $8<z<11$, although they present obvious differences at $k=$ 1 and 3 Mpc$^{-1}$.
This can be understood in terms of heating from XRBs being different from the one of the hot ISM at small scales, while on large scales it is similar.

We note that these variations in the evolution as a function of scale are due to the non-gaussian and very complex processes of ionization and heating from sources with different spectral energy distributions.
We also highlight that the impact of heating and partial ionization on the 21~cm power spectra are difficult to disentangle, as both can on the one hand reduce the amplitude of the signal (by raising the gas temperature and reducing the amount of neutral hydrogen), and on the other increase its fluctuations (through inhomogeneities). Additionally, the impact of the different processes varies with scale and redshift.

\subsection{21 cm Bispectra}

The bispectrum of the 21 cm signal is defined as:
\begin{align}
& b_{\rm 21cm}(\bm{k_1},\bm{k_2},\bm{k_3}) =  \delta_{D}(\bm{k_1}+\bm{k_2}+\bm{k_3} )  \notag\\
&\times \langle \tto(\bm{k_1})\tto(\bm{k_2})\tto(\bm{k_3})\rangle
\end{align}
where $\bm{k_1},\bm{k_2},\bm{k_3}$ are wavenumbers in Fourier space, and $\delta_{D}$ is the Dirac function.
Depending on the values of $\bm{k_1}$, $\bm{k_2}$ and $\bm{k_3}$, i.e. $k_1$, $k_2$ and $k_3$ respectively, the 21 cm bispectra assume different configurations which can probe unique non-gaussian features, e.g. equilateral triangles $k_1=k_2=k_3=k$ (in the following for this case we will only mention $k$ instead of the three $k$s),
isosceles triangles with $k_1=k_2$, and asymmetric triangles with $k_1=n*k_2$, where $n\neq 1$ is a free parameter.
Here we concentrate most of the discussion on the easiest configuration, i.e. the equilateral triangle, which describes the correlation present in the 21~cm signal between three equidistant points in $k$ space.
{The behaviour of such bispectra at different scales can be explained by the skewness once the field has been smoothed over the same scale (see the Appendix A for a detailed discussion of this correspondence).}
The bispectra are computed following the method described in \cite{Watkinson2017}, i.e. we use Fourier transforms, as this is much faster than the traditional method of enumerating triangles, and provides consistent results.
We also normalize $b_{\rm 21cm}({k_1},{k_2},{k_3})$ as $B_{\rm 21cm}(k_1,k_2,k_3) =k_{2}^{3}k_{3}^{3}/(2\pi^{2})^{2} b_{\rm 21cm}(k_1,k_2,k_3)$ \cite[similarly to what done in ][]{Majumdar2020}.

Generally speaking, the sign of the bispectrum can be explained in terms of the concentration of the $\tto$ field, i.e. a positive (negative) bispectrum is obtained when the non-gaussianity of the signal is dominated by an above(below)-average concentration of the DBT field \citep{Hutter2020}. Its shape and evolution are not straightforward to interpret as they depend on the details of the reionization history, but a general behaviour is consistently found by other authors (e.g. \citealt{Majumdar2018}, \citealt{Watkinson2019}, \citealt{Hutter2020}) and confirmed in this work. In the following we discuss it in more detail.

The right panel of Fig.~\ref{fig:s_21cm_ps_vz} shows the redshift evolution of the 21~cm bispectra ($B_{\rm 21cm}$) of equilateral triangles at $k = 0.3$, 1 and 3 Mpc$^{-1}$.
Similarly to what observed for the skewness, at the beginning of reionization, when the fluctuations are dominated by those in the density field\footnote{It should be noted that in this regime the temperature and density fields are correlated, as the former is dominated by the temperature determined by the hydrodynamic simulations.}, i.e. by an above-average concentration, the bispectrum is positive and its amplitude becomes larger as structure formation proceeds and the density concentrations increase.
At $z \gtrsim 11$, thus, the five models display a similar and positive $B_{\rm 21cm}$ at all $k$s.
We note that the temperature from the hydrodynamic simulations becomes higher with decreasing redshift in the vicinity of the sources, thus reducing the amplitude of the signal.
Additionally, as ionized regions continue to form and develop around the sources, although the signal is still dominated by the underlying density and temperature distributions, the amplitude of the bispectrum decreases, because an increasing number of cells have a zero signal and fully ionized regions are correlated to high density regions, so that the non-gaussianity of the signal becomes less dominated by the density peaks.

At $z \lesssim 11$, the effect of energetic sources becomes visible and differences are evident.
With stars only, similarly to what observed for the skewness, a bias appears towards the low temperature tail of the DBT distribution and the bispectrum eventually becomes negative. The timing of the sign transition depends on the scale, i.e. it occurs at $z \sim 10$ for $k = 3 \,\rm Mpc^{-1}$, and at $z \sim 7.5$ for $k= 0.3$ and 1 $\rm Mpc^{-1}$ (we refer the reader to the Appendix A for a more detailed discussion of the sign transition as a function of scale).
When heating of the IGM is very efficient (in the GALISM and GXQI models) most of the gas is in emission and, as observed for the skewness, at small scales the DBT distribution is biased towards the high temperature tail. At larger scales, though, we observe a sign inversion, which can again be understood from the behaviour of the skweness, as, when fluctuations on increasingly large scales are removed, the DBT distribution becomes biased towards its low temperature tail (see Appendix A).
As the XRBs heat less efficiently than the hot ISM, this reduces (increases) the fluctuations at large (small) scales. As a consequence, although the evolution in the GALXRB model is similar to that in the GALISM and GXQI models, it has a lower (higher) absolute amplitude at $k = 0.3$ (3) $\rm Mpc^{-1}$.
The weaker effect is also reflected in the absence of sign transition for $k = 1 \,\rm Mpc^{-1}$.
Because of the rarity of nuclear BHs, the evolution for the GALQSO model is similar to that for the GAL model, although differences are present at $z<9$, e.g. a higher $B_{\rm 21cm}$ at $3\,\rm Mpc^{-1}$, but lower one at $k = 0.3 \,\rm Mpc^{-1}$, and no transitions at $k = 1 \,\rm Mpc^{-1}$.
Towards the end of reionization, the fluctuations in the 21~cm signal are dominated by very concentrated islands of neutral hydrogen, which push again the bispectrum towards positive values and another sign inversion, which is not reached though in all models (see Appendix A).

The right panel of Fig.~\ref{fig:s_21cm_ps_vk} shows the 21 cm bispectra ($B_{\rm 21cm}$) of equilateral triangles in our five models at three redshifts. Its interpretation follows from the discussion above and in the Appendix A.
At $z \gtrsim 11$, when the ionization fraction is very low ($\bar{x}_{\rm HII} = 0.01$), the non-gaussianity of the signal is dominated by that in the density field (and the correlated hydro-temperature from the MBII simulations), and thus the five models have a positive $B_{\rm 21cm}$.
Similarly to the power spectra shown in the left panel, the 21 cm bispectra display obvious features associated to the presence of X-ray sources.
As the heating of energetic photons washes out the small scale fluctuations while increasing the large scale ones, the 21 cm bispectrum of models including sources other than stars is larger than those of GAL and GALQSO at $k \lesssim 1\,\rm Mpc^{-1}$, but lower at $k \gtrsim 1\,\rm Mpc^{-1}$.
At $z=8$, when the hot-ISM has already partially ionized and heated up the IGM, the 21 cm bispectrum of GALISM and GXQI models is negative at $k<1.2\,\rm Mpc^{-1}$.
As the harder photons emitted by XRBs are less efficient at heating, this results in a 21~cm bispectrum of the GALXRB model which is negative only at $k<0.6\,\rm Mpc^{-1}$.
The $B_{\rm 21cm}$ of GAL and GALQSO models are similar and positive at $k<2\,\rm Mpc^{-1}$, but negative at $k\sim 3\,\rm Mpc^{-1}$.
Their opposite sign compared to the other three models is due to the lack of heating and partial ionization from energetic photons, which results in many cells with a negative $\tto$ (see the Fig.~\ref{fig:s_21cm_image} and Fig.~\ref{fig:s21cm_1d_ts}).
At $z =7$, the bispectra of the GALXRB, GALISM and GXQI models are similar to those at $z=8$, although the transition from positive to negative sign has shifted to larger scales.
The rare but very luminous nuclear BHs have obvious contributions at $z=7$, but, as their ionization and heating are not as uniform as those induced by ISM and XRBs, the bispectrum of the GALQSO model is more complex, with a positive sign at $k=[0.8-1.6]\,\rm Mpc^{-1}$ and $k>4 \,\rm Mpc^{-1}$, and a negative one otherwise.
The 21~cm bispectrum of GAL model is negative in the full $k$ range studied, since most neutral cells in this model have $\tto<0$~mK (see the right panel of Fig.~\ref{fig:s21cm_1d_ts}).

\begin{figure}
    \centering
    \includegraphics[width=0.95\linewidth]{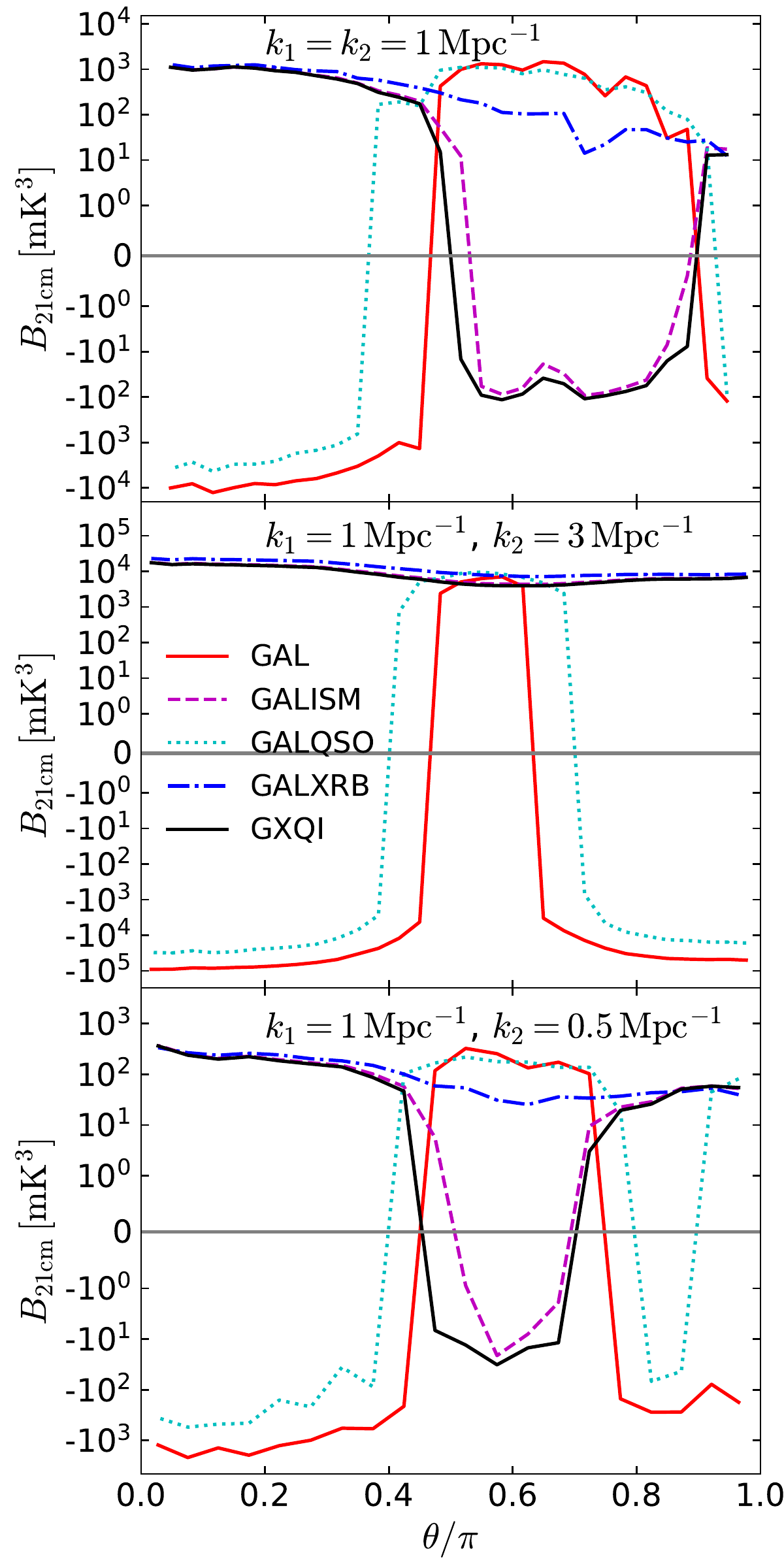}
    \caption{Bispectra of non-equilateral triangles of 21 cm signal in models GAL (solid red), GALISM (dashed magenta), GALQSO (dotted cyan), GALXRB (dash-dotted blue) and GXQI (solid black) at $z=8$ ($\bar{x}_{\rm HII} = 0.21$), as a function of angle $\theta/\pi$ between $k_{1}$ and $k_{2}$.
    From top to bottom, the panels refer to $k_{1}=k_{2}=1\,\rm Mpc^{-1}$, $k_{2}=3k_{1}=3\,\rm Mpc^{-1}$ and $k_{2}=0.5k_{1}=0.5\,\rm Mpc^{-1}$.
    The horizontal gray lines are drawn at zero to guide the eye.}
    \label{fig:bispec_21cm_kneq3_vk}
\end{figure}
In addition to the equilateral triangles, the 21~cm bispectra have many other independent modes \citep{Bharadwaj2020,Majumdar2020} which might be able as well to highlight the differences between various source models. While we will not show all cases, we will discuss at $z=8$ ($\bar{x}_{\rm HII} = 0.21$) some representative cases with a fixed $k_{1}=1\,\rm Mpc^{-1}$ and three different values of $k_{2}$,  i.e. isosceles triangles, namely $k_{2}=k_{1}=1\,\rm Mpc^{-1}$, a larger $k_{2}=3 k_{1}=3\,\rm Mpc^{-1}$ and a smaller $k_{2}=0.5 k_{1}=0.5\,\rm Mpc^{-1}$.
$k_{3}$ is a free parameter varying in the range $|k_{1}-k_{2}| \le k_{3} \le |k_{1}+k_{2}|$), so, instead of $k_{3}$, we show the results in terms of the angle $\theta$ between $k_{1}$ and $k_{2}$, which is computed as:
\begin{equation}
\theta =\pi -{\rm arccos}\left(\frac{k_{1}^{2}+k_{2}^{2}- k_{3}^{2}}{2k_{1}k_{2}}\right).
\end{equation}
The 21~cm bispectra of non-equilateral triangles in the five models are shown in Fig.~\ref{fig:bispec_21cm_kneq3_vk}.
As the hot ISM already fully heats up the IGM at $z=8$, $B_{\rm 21cm}$ in the GALISM and GXQI models is dominated by the ionization and the matter density components.
When $k_{2}/k_{1} \gg 1$, e.g. for $k_{2}=3k_{1}=3\,\rm Mpc^{-1}$, $B_{\rm 21cm}$ is dominated by the matter density and thus positive in the whole $\theta$ range.
When $k_{2}/k_{1} \sim 1$, e.g. for $k_{2}=k_{1}=1\,\rm Mpc^{-1}$ and $k_{2}=0.5 k_{1}=0.5\,\rm Mpc^{-1}$, $B_{\rm 21cm}$ is dominated by matter at small and large $\theta$ but by ionization in the middle, resulting in a negative $B_{\rm 21cm}$ for $\theta=[0.5-0.9]\pi$ and  $k_{2}=k_{1}=1\,\rm Mpc^{-1}$, and for $\theta = [0.4-0.7]\pi$ and $k_{2}=0.5 k_{1}=0.5\,\rm Mpc^{-1}$.
This behaviour is extensively discussed in \cite{Majumdar2018} and we consistently confirm their results.
{As the XRBs do not fully heat up the IGM at $z=8$, the effect of ionization resulting in a negative $B_{\rm 21cm}$ in the GXQI and GALISM models is reduced, so that $B_{\rm 21cm}$ in the GALXRB model remains positive at all scales considered.}
Similarly to the $B_{\rm 21cm}$ of equilateral triangles shown in Fig.~\ref{fig:s_21cm_ps_vk}, those of non-equilateral triangles in the GAL and GALQSO models have a sign opposite to the one of the GALISM and GXQI models.
This is because the former models have mostly neutral cells with a negative $\tto$.
There is an exception at $\theta \sim 0.55$ in the case of $k_{2}=3 k_{1}=3\,\rm Mpc^{-1}$, since the weak heating of hard UV photons from stellar sources can also heat up the IGM and result in a behaviour similar to that of the  models including energetic sources.
Although it is still very weak at $z=8$, the contribution of nuclear BHs produces differences in the $B_{\rm 21cm}$ of non-equilateral triangles compared to the GAL model, which are even more significant than those of equilateral triangles.
This means that in some cases the 21 cm bispectra of non-equilateral triangles can be a tool more powerful than those of equilateral triangles to distinguish the differences between various source models.
To investigate in even greater detail the capability of 21 cm bispectra to study the EoR, one can adopt the method presented in  \cite{Bharadwaj2020} and \cite{Majumdar2020} to show all the modes at $k_{1}$s. This is though beyond the scope of the present paper.

\subsection{Detectability}
In the near future, the 21 cm power spectra and bispectra are expected to be measured by 21~cm facilities such as LOFAR, MWA, SKA1-low and HERA \citep{Yoshiura2015, Watkinson2019}.
In the following, we will use SKA1-low as our facility of reference to study the capability of 21~cm experiments to disentangle the impact that different source models have on the IGM properties and thus on the 21~cm power spectra and bispectra.
Although also LOFAR, MWA, and HERA might be able to measure the 21~cm power spectra and bispectra \cite[e.g. ][]{Shimabukuro2017,Thyagarajan2020}, as they reach the maximum sensitivity for an angular resolution which is not covered by the box size of our simulations, we do not present any results for these telescopes.

SKA1-low is designed to have 224 stations in a compact core with a diameter of 1~km, and 224 stations in three arms with a baseline that can extend up to 65~km\footnote{https://astronomers.skatelescope.org/}.
Its frequency range, 50--350 MHz, covers the 21~cm signal from $z=3$ to 27.
The 21 cm power spectra and bispectra measurements by SKA1-low are mainly from the compact core, with an angular resolution $\vartheta = \lambda/D$, where $D=1\,\rm km$ is the largest distance between two stations.
This corresponds to a $k \sim 0.3\,\rm Mpc^{-1}$ at $z \sim 9$.
The rms of noise brightness temperature can be simply estimated as:
\begin{equation}
    T_{\rm N} =  \frac{\lambda^{2} T_{\rm sys}}{A_{\rm eff} \Omega_{\rm beam} \sqrt{N_{\rm st}(N_{\rm st}-1) B_{\rm width} t_{\rm int}}},
\end{equation}
where $T_{\rm sys}$ is the system temperature, $A_{\rm eff}$ is the effective collecting area, $N_{\rm st}=224$ is the number of stations inside the core, $B_{\rm width} = 0.1 \,\rm MHz$ is the spectral resolution adopted, $t_{\rm int} = 1000 \,\rm hours$ is the integration time, and the solid angle of one measured pixel is $\Omega_{\rm beam} = 1.133\vartheta^{2}$.
The sensitivity of the stations $S =A_{\rm eff} / T_{\rm sys}$ is taken from \cite{Dewdney2016}.
Here we simply assume that all the station pairs have the same angular resolution, i.e. they have the same distance $D$.
Note that an accurate calculation of the SKA1-low noise spectrum would require a realistic distribution of the antennas and the simulation of the uv-coverage \citep{Haarlem2013, Dewdney2016}.

Assuming the instrumental noise is completely gaussian, the power spectrum of noise can be estimated as $P_{\rm N} = T_{\rm N}^{2} x^{2}y$ \cite[for a more accurate computation, please refer to e.g.][]{Yoshiura2015},
where $x$ and $y$ are the comoving length corresponding to the angular and frequency resolution, respectively.
Including the sampling error, the expected error on measured power spectra is
\begin{equation}
\sigma_{\rm N} =\sqrt{(P_{\rm 21cm}^{2} + P_{\rm N}^{2})/(0.5N_{\rm pair})},
\end{equation}
where $N_{\rm pair} =4\pi k^{2} {\rm d}k V/(2\pi)^{3}$, ${\rm d}k = 0.23 k$ (i.e. ${\rm d} \, {\rm log}_{10}(k) = 0.1$) is the $k$ bin-width, $V$ is the comoving volume covered by the field of view of SKA1-low and the frequency bandwidth at $z$.
Here, we take a frequency bandwidth of $2\,\rm MHz$.
Considering that only half of the $k$ modes after Fourier transform are independent, we multiply $N_{\rm pair}$ by 0.5.
While white noise has no bispectrum, as the latter describes non-gaussian features, it nevertheless pollutes measured bispectra with statistical noise, with an amplitude that can be estimated as
$b_{\rm N} = T_{\rm N}^{3} x^{4}y^{2}$ \citep{Yoshiura2015}.
Considering the sampling error, the total bispectrum error is
\begin{equation}
\Sigma_{\rm N} =\sqrt{(b_{\rm 21cm}^{2} + b_{\rm N}^{2})/(N_{\rm tri}/12)},
\end{equation}
where $N_{\rm tri}$ is the number of triangles in the comoving volume $V$.
As only two $\bm{k}$s are free to configure equilateral triangles, $N_{\rm tri}$ can be estimated as:
\begin{equation}
N_{\rm tri} =\left[\frac{4\pi k^{2} {\rm d}k V}{(2\pi)^{3}}\right] \times  \left[\frac{2\pi {\rm sin}(2\pi/3){\rm d}\alpha k^{2} {\rm d}k V}{(2\pi)^{3}}\right],
\end{equation}
where the first factor represents the sum of ${\bm k}_{1}$ in the $k$ bin-width, and the second the sum of ${\bm k}_{2}$.
For a fixed ${\bm k}_{1}$, ${\bm k}_{2}$ comprises only those with an angle $2\pi/3$ respect to ${\bm k}_{1}$ within a bin-width ${\rm d}\alpha$.
Here we set ${\rm d}\alpha$ consistently with the resolution in Fourier space.
As one triangle is repeated 6 times, and again only half of the $k$ modes are independent, we divide  $N_{\rm tri}$ by a factor of 12.
We refer the reader to \cite{Yoshiura2015} for more theoretical details about the estimation of power spectra and bispectra noise.
Finally, we assume that foreground contamination can be removed without any residual (see e.g. \citealt{Geil2008}).

\begin{figure*}
    \centering
    \subfigure{
    \begin{minipage}{0.45\linewidth}
    \centering
    \includegraphics[width=1.0\linewidth]{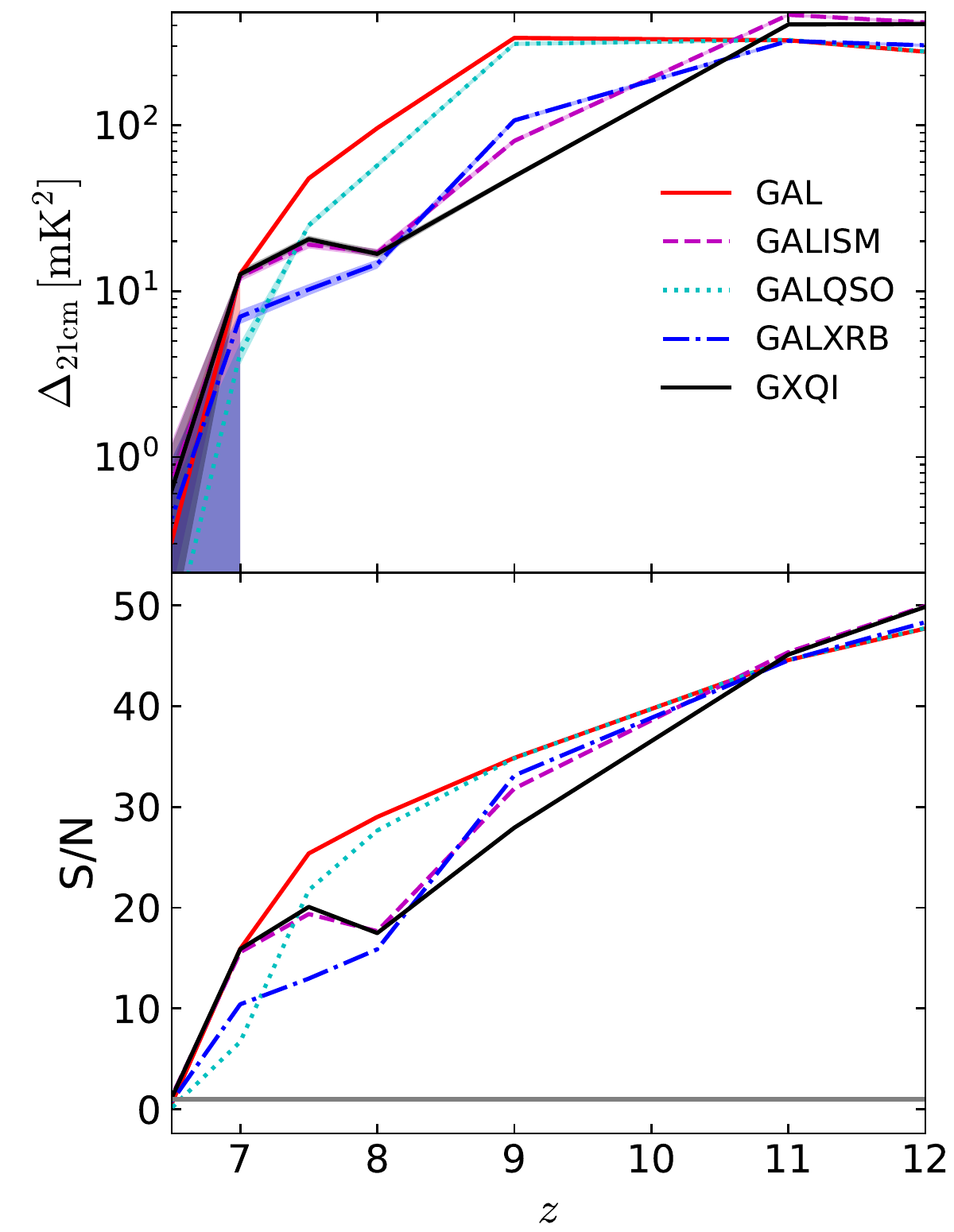}
    \end{minipage}
    }
    \subfigure{
    \begin{minipage}{0.45\linewidth}
    \centering
    \includegraphics[width=1.0\linewidth]{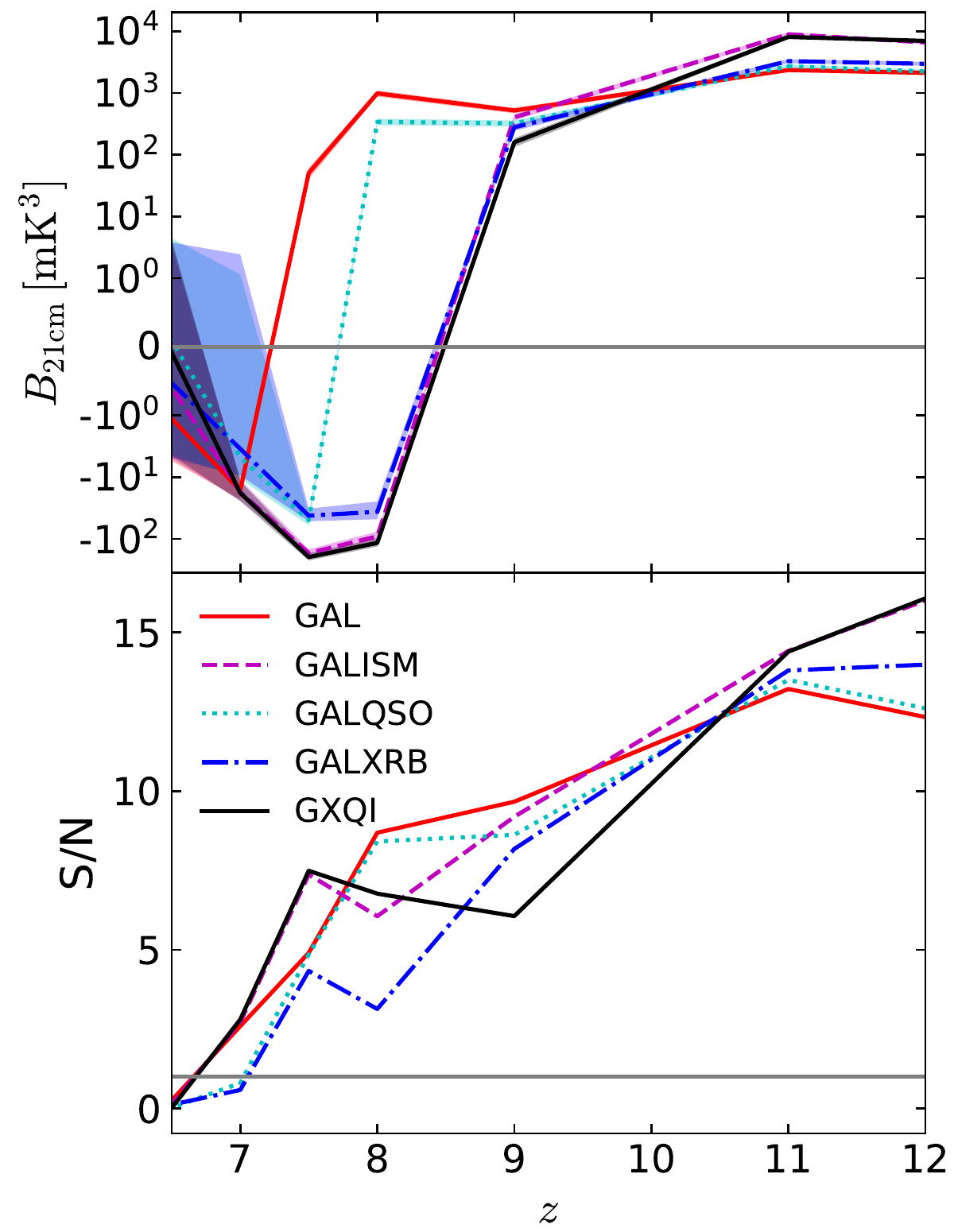}
    \end{minipage}
    }
    \caption{Predicted 1-$\sigma$ ranges (top panels)  and S/N ratios (bottom) of  21 cm power spectra (left) and bispectra (right) of equilateral triangles in model GAL (solid red), GALISM (dashed magenta), GALQSO (dotted cyan), GALXRB (dash-dotted blue) and GXQI (solid black) at $k=0.3\,\rm Mpc^{-1}$ as a function of redshift.
     The horizontal gray lines in the bottom panels denote S/N $=1$, the one in the top right panel is the zero line to guide the eye.
     }
    \label{fig:detec_21cm}
\end{figure*}
Fig.~\ref{fig:detec_21cm} shows the estimated 1-$\sigma$ regions and S/N ratios of 21 cm power spectra and bispectra of equilateral triangles at $k=0.3\,\rm Mpc^{-1}$ in our five models from $z=6.5$ to 12.
Although the instrumental noise increase quickly with increasing $z$ \citep{Dewdney2016}, this is true also for the $P_{\rm 21cm}$ and $B_{\rm 21cm}$, so that their expected S/N ratios become higher at earlier times (see also e.g. \citealt{Pacucci2014} and \citealt{Fialkov2017} for similar results).
These are much larger than 1, except at the end of the EoR when most of the IGM is in a highly ionization state, so that SKA1-low should be easily able to measure both $P_{\rm 21cm}$ and $B_{\rm 21cm}$, and possibly to distinguish the features caused by sources more energetic than stars.
More specifically, the 1-$\sigma$ noise on the 21~cm power spectra is larger than the differences between our five models at $z<7$, while they are much smaller at $z>7$. For example, the GALISM and GXQI models have a similar $P_{\rm 21cm}$, but their differences are still larger than the noise at $z \sim 9$.
The S/N ratios of the 21 cm power spectra are larger than 15, 25 and 40 at $z=8$, 9 and 11, respectively. At even higher redshifts, when the effect of X-ray sources is not very strong, the power spectra are very similar and thus it becomes more challenging to distinguish the various scenarios.

Although the errors on the bispectra are relatively bigger, resulting in lower S/N ratios (though still larger than 3, 6 and 13 at $z=8$, 9 and 11), SKA1-low should still be able to distinguish differences between the five models at $z>7.5$, especially at $z\sim 8$, while this becomes increasingly difficult at higher $z$, where the bispectra from different models have similar values.
Note that, as for the power spectra, the bispectra can be heavily polluted by the foreground noise \citep{Watkinson2020}, significantly reducing the S/N.
Finally, while here we show only one example of bispectra, a combination of different modes should provide further insight into the relative role of X-ray sources during the EoR.

\section{Discussion and Conclusion}
\label{sec:con}

We have studied various 21 cm statistics associated to X-ray source models during the epoch of reionization (EoR), e.g. global mean, deviation, skewness, power spectrum and bispectrum, using the cosmic reionization histories \citep{Eide2018, Eide2020} obtained from the high resolution cosmological hydrodynamical simulation MBII \citep{Khandai2015} post-processed with the 3-D multi-frequency radiative transfer (RT) code CRASH (e.g. \citealt{Ciardi2001,Maselli2009,Graziani2018}).
This is the first study that is based on systematic simulations of reionization with RT that captures accurately the heating and ionization of various energetic sources modelled with hydrodynamic simulations.
We have analysed five RT simulations with the same stellar ionization model, while the contribution from more energetic sources differs. More specifically, the reference simulation has only stars (GAL), while the others additionally have hot ISM (GALISM), accreting nuclear BHs (GALQSO), XRBs (GALXRB), and all sources combined (GXQI).
These X-ray sources have a negligible contribution to the global ionization history, while they strongly affect the gas partial ionization and temperature, and, consequently, the 21 cm signal.
We note that the emissivity of the X-ray sources in our simulations are consistent with measurements of the global X-ray background and the angular power spectrum of its fluctuations \cite[see][who adopted the same simulations analysed here]{Ma2018b}.
More specifically, the XRBs contribute a few percents to the background and fluctuations, the nuclear BHs $\sim 25\%$ of the XRBs to the background, while the contribution of the hot ISM is negligible.
This suggests that our models are conservative in terms of emission (and thus heating) from X-ray sources.
Thus, while our source modeling relies entirely on the hydrodynamical simulations, different prescriptions (in particular for what concerns the BHs; in this respect we refer the reader also to \citealt{Eide2020b}) can increase the impact of X-ray sources without violating existing observational constraints. Additional constraints on such sources can come in the future by e.g. observations of the cross-correlation between the X-ray background and the 21 cm signal (see e.g. \citealt{Ma2018b}).

The XRBs with spectra from \cite{Fragos2013a,Fragos2013b} are inefficient at heating the neutral gas \citep{Fialkov2014,Eide2020}, so that many cells in the GALXRB model still have $\tto<0$~mK at $z=8$ ($\bar{x}_{\rm HII} = 0.21$). Similarly, although nuclear BHs are locally much more effective, due to their paucity they have a negligible global effect, so that the behaviour of the GALQSO simulation is very similar to the reference one, although at $z \lesssim 9$ some small differences become visible. We note that the impact of the nuclear BHs here is much weaker than in e.g. \cite{Ross2019}, since the latter have a higher number density of accreting BHs.
On the other hand, the hot ISM is very efficient at uniformly heating the IGM, so that in the GALISM and GXQI models, the 21 cm power spectrum is $\Delta_{\rm 21 cm} \sim 10 \si{\mK}^{2}$ at $z=7$ ($\bar{x}_{\rm HI} = 0.76$), consistent with other values quoted in the literature \cite[e.g. ][]{Shaw2020, Mondal2015}.
Our power spectra at $z>10$, though, are typically higher than those in the literature, since the X-ray sources in our models are not efficient at heating at high-$z$.
Note, though, that we assume that the spin temperature is coupled to the kinetic temperature through Ly$\alpha$ scattering. Although a detailed evaluation of the Ly$\alpha$ background is beyond the scope of this paper, we have estimated that it should become higher than the threshold value needed for an efficient coupling at $z \lesssim 13$, i.e. the results of 21 cm power spectra and bispectra at $z>13$ might be overestimated.

Compared to the power spectrum, the 21 cm bispectrum is more sensitive to the ionization and heating process.
On some scales its sign is expected to be negative when hydrogen is moderately ionized, and positive when it is highly ionized \citep{Majumdar2018, Hutter2020} or little ionized but warm \citep{Watkinson2019}.
Our results qualitatively confirm these conclusions and also more quantitatively the amplitude of the bispectra are e.g. consistent with those of \cite{Majumdar2020}.
We find that some of the characteristics of the bispectra, e.g. the transition between a positive and negative sign, are very sensitive to the properties of the X-ray sources included in the models, so that even towards the end of reionization the impact of sources which are very efficient at uniformly heating the IGM (such as the hot ISM) can be clearly distinguished from that of the less efficient XRBs and nuclear BHs.

{Finally, we note that while the source emission characteristics have been derived directly from the physical properties of stars, galaxies and BHs modelled in the MBII simulations (i.e. we have not included any additional parameter with the exception of the escape fraction), the same 21~cm signal could be obtained with a different combination of properties.
Such degeneracy poses a real challenge to an unambiguous  determination of the relative contribution to reionization of different source types. While faster modeling algorithms are employed to investigate a large parameter space (see e.g. \citealt{Ghara2020}), here we note that combining observations at multiple redshifts and scales, and exploiting the fact that the various sources behave differently e.g. in terms of transitioning from a positive to a negative global signal and bispectrum, should allow in the future to constrain the spectral energy distribution of the sources which contribute to the reionization and reheating process.}

Our main results can be summarized as follows:
\begin{itemize}
\item The GAL model has a negative $\meantto$ throughout the whole EoR, because the short mean free path of the UV photons emitted by the stars does not allow for heating of the neutral gas outside of the fully ionized regions. For the same reason, this model has also the largest deviation $\sigma_{\rm 21cm}$. Its skewness is positive at the highest redshift and changes sign at $z \sim 9$.
Due to the negligible effect of nuclear BHs on the global reionization process, the GALQSO model has a $\meantto$ and a $\sigma_{\rm 21cm}$ which are similar to those of the GAL model, although the skewness at $z \lesssim 8$ deviates from that of the GAL model and becomes positive again at $z \sim 7$, when the heating from the BHs is more relevant.
Due to the effect of heating from energetic photons, the GALXRB, GALISM and GXQI models show similar results, with a transition from negative to positive $\meantto$ which happens earlier for the models with more effective heating, i.e. first for GXQI, followed by GALISM and GALXRB. This results also in a higher (lower) deviation for the GALXRB (GXQI) model. The efficient heating is also responsible for a positive skewness during the whole EoR.

\item Because of the absence of heating of the neutral hydrogen, the power spectrum of the GAL model is higher than all the others at almost all redshifts and scales. The maximum difference is reached at $z \sim 9$, when it is about 10 times higher than the power spectrum of the GXQI model. The power spectrum in the GALQSO model follows closely the GAL one, although at $z \lesssim 9$ the impact of the BHs can be seen in terms of an amplitude reduction.
The partial ionization and heating of the other energetic sources (in particular the hot ISM) reduces the amplitude of the power spectra at $z \lesssim 13$, while it also increases it on large scales ($k \lesssim 0.5\,\rm Mpc^{-1}$) at $z \gtrsim 11$.

\item While all models have positive and similar bispectra (of equilateral triangles) at high redshift, the ionization process induces a transition to negative values on large scales, which is obviously affected by the presence of energetic sources.
For example, it happens at $k< 1\,\rm Mpc^{-1}$ and $z \sim 8$ in the GALISM and GXQI models, while at a lower $k$ in the GALXRB model due to the weaker heating and partial ionization of the XRBs. The transition is delayed even further for the GAL and GALQSO models. On the smaller scales (i.e. $k \sim 3\,\rm Mpc^{-1}$) though, the bispectra of the GALXRB, GALISM and GXQI models remains always positive.
The bispectra of non-equilateral triangles also show obvious differences between the five models.

\item The SKA1-low is expected to measure the 21 cm power spectra and bispectra for all five models with high S/N ratios. At $z=8$, 9 and 11, and $k = 0.3\,\rm Mpc^{-1}$  these can reach values larger than $15$, 25 and 40 respectively for the power spectra and $3$, 6 and 13 for the bispectra.
\end{itemize}

We conclude by noting that the next generation of radio telescopes is expected to measure with high S/N ratios various statistics associated to the 21~cm signal from the EoR. Our systematic investigation of the impact of different source types shows that such observations should also be able to distinguish between the various sources, as they leave a clear imprint on the different statistics.

\acknowledgments

The authors thank an anonimous referee for her/his useful comments.
BC is grateful to Eichiiro Komatsu for an enlightening discussion on bispectra.
The tools for bibliographic research are offered by the NASA Astrophysics Data Systems and by the JSTOR archive.
QM is supported by National Natural Science Foundation of China (Grant No. 11903010), innovation and entrepreneurial project of Guizhou province for high-level overseas talents (Grant No. (2019)02), Science and Technology Fund of Guizhou Province (Grant No. [2020]1Y020), GZNU 2018 doctoral research funding (Grant No. GZNUD[2018]9) and GZNU 2019 Special project of training new academics and innovation exploration.
YM is supported by the National Key R\&D Program of China (Grant No. 2018YFA0404502, 2017YFB0203302),
and the National Natural Science Foundation of China (NSFC Grant No. 11673014, 11761141012, 11821303).
QZ is supported by National Natural Science Foundation of China (Grant No. U1731238) and Foundation of Guizhou Provincial Education Department (No. KY[2020]003).

\appendix
\label{app}
\section{Explaining the evolution of the 21 cm bispectrum using smoothed skewness}

As discussed in e.g. \cite{Shimabukuro2016}, the skewness is related to the bispectrum by:
\begin{equation}
\label{eq:skew_bis}
    \skewtto = \frac{1}{\left(\stdto\right)^3} \int \frac{{\rm d}^{3} k_{1}}{(2\pi)^{3}} \int \frac{{\rm d}^{3} k_{2}}{(2\pi)^{3}} b_{\rm 21cm}(\bm{k_{1}}, \bm{k_{2}}, -\bm{k_{1}}-\bm{k_{2}}),
\end{equation}
suggesting that the evolution of the bispectrum can be described in terms of the skewness.
However, the latter is the integration of all the bispectrum modes, and thus one 21 cm field has only one skewness, as shown in the left panel of Fig.~\ref{fig:s21cm_1d_ts}.
By assuming that the 21 cm bispectra are smooth with $k$, the integration of Eq.~\ref{eq:skew_bis} over $k<k_{\rm max}$ (i.e. the skewness of the 21~cm signal when scales with $k>k_{\rm max}$ are smoothed) should represent the behaviour of the bispectrum at $k \sim k_{\rm max}$, due the amplitude of the bispectrum being higher at large than at small $k$s (see e.g.  Fig.~\ref{fig:s_21cm_ps_vk}) and to a significantly larger number of small scale modes than large scale ones.
Thus, in this appendix we try to explain the behaviour of the 21 cm bispectrum at different $k$s with 21 cm skewness of fields after removing the small scale fluctuations.
Since the skewness describes the bias of a field compared to its mean value, this is consistent with the explanation in \cite{Hutter2020} of the sign of the bispectrum being positive/negative when the non-gaussianity of the signal is dominated by above/below average values.

\begin{figure*}
    \centering
    \includegraphics[width=0.95\linewidth]{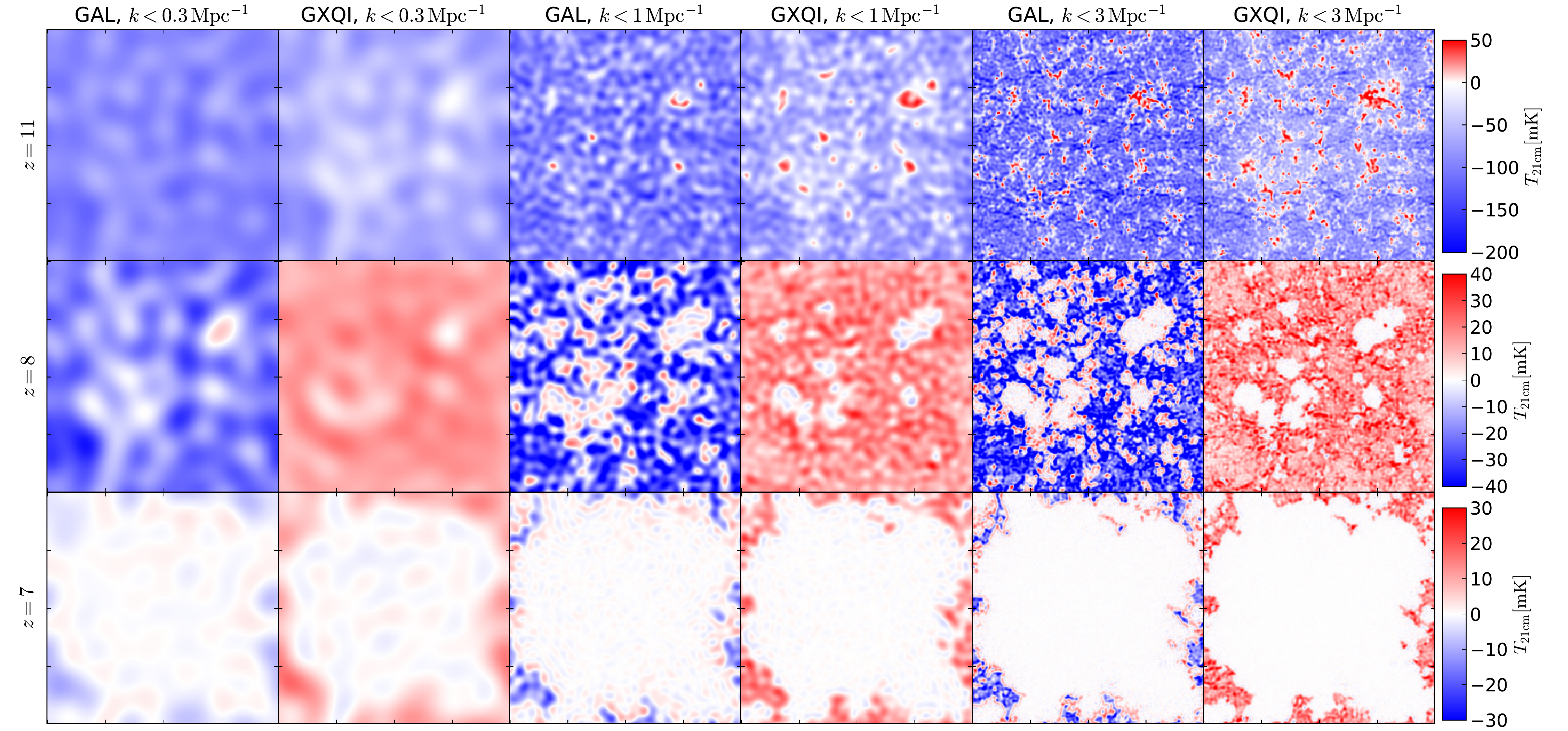}
    \caption{Slices of thickness $0.4h^{-1}\mathrm{cMpc}$ and side length 100 $h^{-1}$ cMpc showing the 21 cm differential brightness temperature in models GAL and GXQI with fluctuations at $k<0.3\,\rm Mpc^{-1}$, $1\,\rm Mpc^{-1}$ and $3\,\rm Mpc^{-1}$ (see labels), at $z = 11$ (top panels, $\bar{x}_{\rm HII} = 0.01$), $z = 8$ (middle, $\bar{x}_{\rm HII} = 0.21$) and $z = 7$ (bottom, $\bar{x}_{\rm HII} = 0.76$).}
    \label{fig:s_21cm_image_kscut}
\end{figure*}
\begin{figure}
        \centering
        \includegraphics[width=0.95\linewidth]{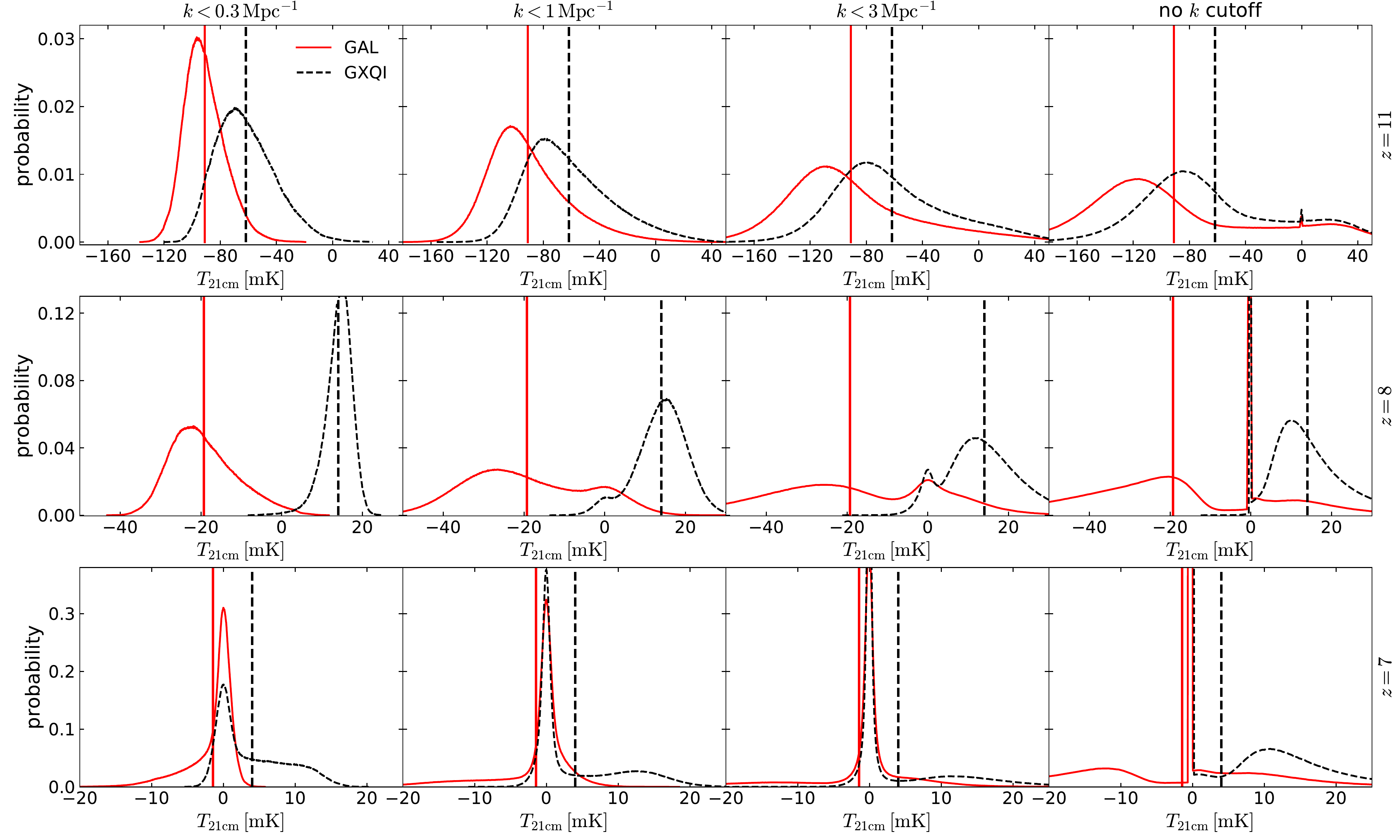}
    \caption{Probability density distributions of $\tto$ in the model GAL (solid red lines) and GXQI (dashed black) with fluctuations at $k<0.3\,\rm Mpc^{-1}$, $1\,\rm Mpc^{-1}$ and $3\,\rm Mpc^{-1}$ from left to right,  at $z = 11$ ($\bar{x}_{\rm HII} = 0.01$), 8 ($\bar{x}_{\rm HII} = 0.21$) and 7 ($\bar{x}_{\rm HII} = 0.76$) from top to bottom. The last column refers to the case in which all fluctuations are retained.
    The vertical lines denote the mean values of $\tto$.
    }
    \label{fig:s_21cm_ts_kscut}
\end{figure}

As an example, in Fig.~\ref{fig:s_21cm_image_kscut} we show maps of differential brightness temperature for the GAL and GXQI models after removing small scale fluctuations, i.e. only fluctuations on scales $k<0.3\,\rm Mpc^{-1}$, $1\,\rm Mpc^{-1}$ and $3\,\rm Mpc^{-1}$ are retained.
The corresponding probability density distributions are shown in Fig.~\ref{fig:s_21cm_ts_kscut}, where, for comparison, we also plot the curves without smoothing, i.e. the ones in the right panel of Fig.~\ref{fig:s21cm_1d_ts}.
Note that here we do not exclude the fully ionized cells.
\begin{table}
    \centering
    \begin{tabular}{c|c|c|c|c|c}
         \hline
         model & $z$ &  $k<0.3\,\rm Mpc^{-1}$ & $k<1\,\rm Mpc^{-1}$ & $k<3\,\rm Mpc^{-1}$ & no $k$ cutoff \\
         \hline
         GAL & 11    &  0.57 & 0.82 & 0.77  & 0.57 \\
         GXQI &  11  &  0.54 & 0.69 & 0.58  & 0.46 \\
         GAL & 8     &  0.55 & 0.17 & -0.22 & -0.47\\
         GXQI &  8   & -1.02 &-0.40 & 0.60  & 1.71 \\
         GAL & 7     & -1.81 &-1.94 & -1.82 & -1.47\\
         GXQI &  7   & 0.72  & 1.17 & 1.83  & 2.75 \\
         \hline
    \end{tabular}
    \caption{Skewness of the 21 cm differential brightness temperature in the GAL and GXQI models at $z=11$, 8 and 7, with and without small scale fluctuations removal.}
    \label{tab:skewness_kscut}
\end{table}

At $z=11$, in both models and at all scales, the mean values lie to the right of the distribution peaks, i.e. the cells with high $\tto$ dominate the skewness and give the  positive values in Table~\ref{tab:skewness_kscut}. This is consistent with the positive bispectrum shown in Fig.~\ref{fig:s_21cm_ps_vk} and Fig.~\ref{fig:s_21cm_ps_vz}.
At $z=8$, with no smoothing the mean value is to the right/left of the peak in the GXQI/GAL model, resulting in a positive/negative skewness. This means that the non-gaussianity of the signal is dominated by the partially ionized and hot gas in the presence of energetic sources, and by the cold neutral gas with only stars.
Smoothing fluctuations at  $k>3\,\rm Mpc^{-1}$ does not change the sign of the skewness, although it reduces its amplitude, as can also be seen from the maps in Fig.~\ref{fig:s_21cm_image_kscut}.
With a suppression of fluctuations on larger scales, i.e. with $k<0.3$ and 1 $\,\rm Mpc^{-1}$, we observe an inverted behaviour, with the mean value on the left/right of the peak in the GXQI/GAL model, resulting in a negative/positive skewness. This sign inversion is clearly observed also in the bispectra at $k=0.3$ and 1 $\rm Mpc^{-1}$ in Fig.~\ref{fig:s_21cm_ps_vk} and Fig.~\ref{fig:s_21cm_ps_vz}.
At $z = 7$, when most cells are ionized, the mean values of $\tto$ are close to $0\,\rm mK$. Then, the few neutral and warm cells with a high value of $\tto$ in the GXQI model bias the skewness towards positive values, while the opposite is true for the GAL model, where the remaining neutral cells are cold. This behaviour is again consistent with the one of the bispectrum in Fig.~\ref{fig:s_21cm_ps_vk} and Fig.~\ref{fig:s_21cm_ps_vz}, with the exception of the scale $k = 0.3 \rm Mpc^{-1}$, where in the GXQI model the bispectrum is negative. We observe, though, that the skewness for $k < 0.3 \rm Mpc^{-1}$ is only 0.72, suggesting that the computation of the bispectrum at large scales is affected by a large error.

{In summary, due to the non-gaussian nature of the reionization process, the probability density distribution of $\tto$ after smoothing of small scale fluctuations (i.e. only keeping $k<k_{\rm max}$) shows a dependence on the upper limit $k_{\rm max}$.
When the distribution peak is on the left of the mean value, it results in a positive skewness and thus a positive bispectrum, while a negative skewness and bispectrum appear when the distribution peak is on the right of the mean value.
Such relation can help to understand why the 21 cm bispectrum is negative or positive at specific $k$s and $z$s.}

\section{Effects of correction for partially ionized cells containing ionization fronts}
{
The effect of correcting the physical state of cells containing the ionization front on the power spectra and bispectra depends on the source models, redshift and wavelength $k$s.
\begin{figure*}
    \centering
    \subfigure{
    \begin{minipage}{0.45\linewidth}
    \centering
    \includegraphics[width=1.0\linewidth]{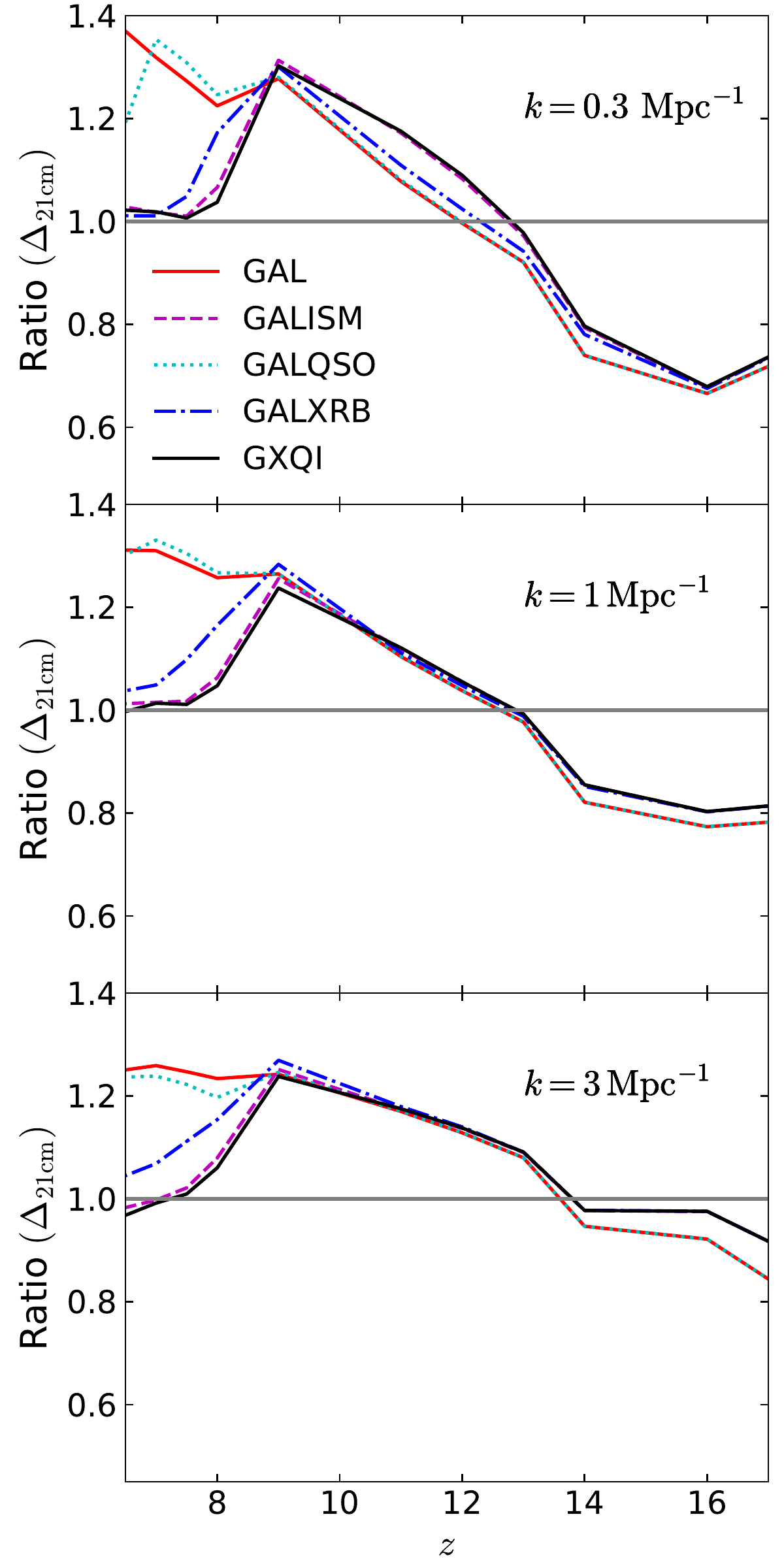}
    \end{minipage}
    }
    \subfigure{
    \begin{minipage}{0.45\linewidth}
    \centering
    \includegraphics[width=1.0\linewidth]{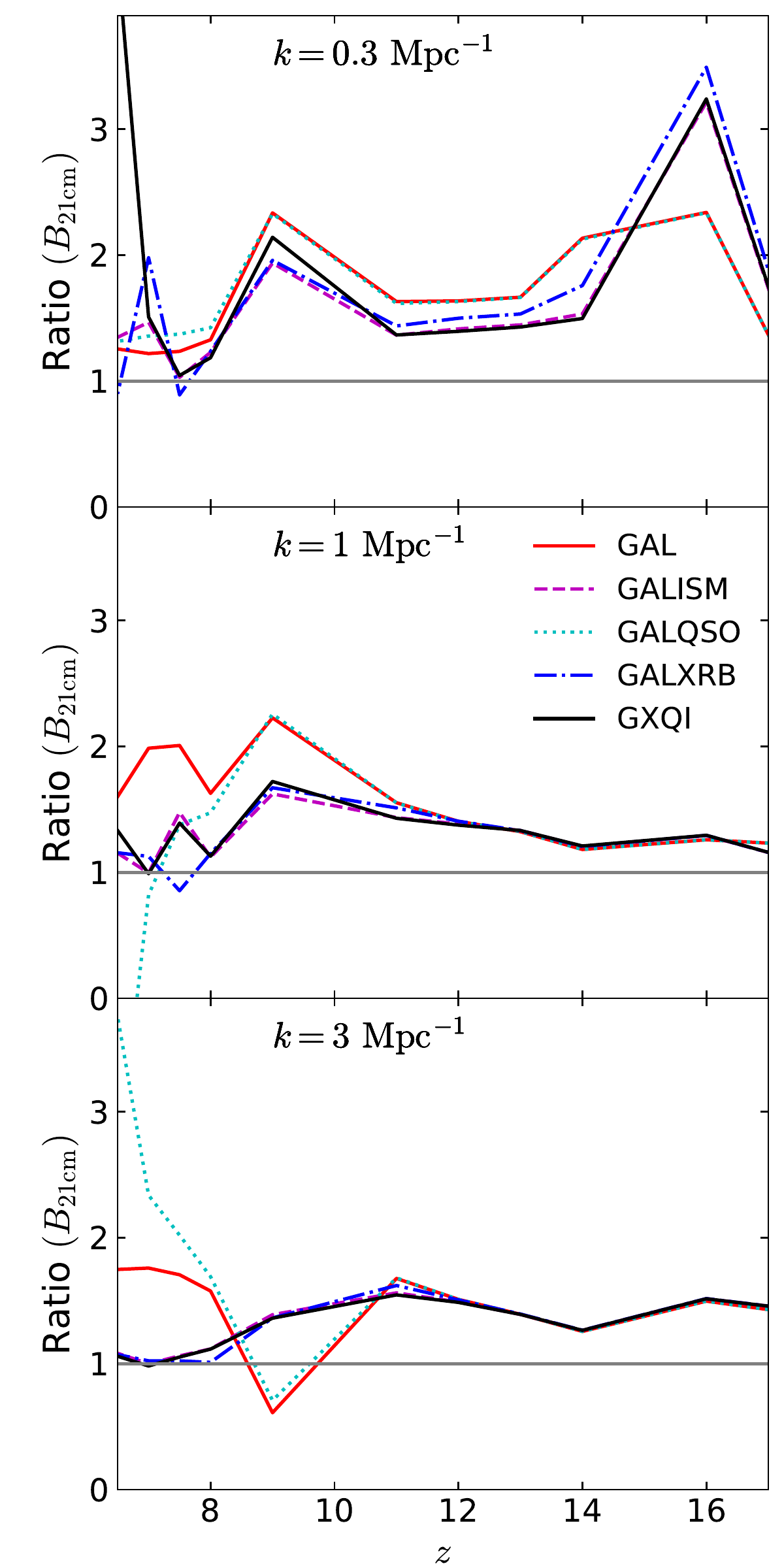}
    \end{minipage}
    }
    \caption{Ratios between 21 cm power spectra (left) and bispectra of equilateral triangles (right) without and with correction in model GAL (solid red), GALISM (dashed magenta), GALQSO (dotted cyan), GALXRB (dash-dotted blue) and GXQI (solid black) at $k=0.3$ (top), 1 (central) and 3 (bottom) $\rm Mpc^{-1}$ as functions of redshift. The horizontal gray lines denote ratios $=1$ to guide the eye.
    }
    \label{fig:psbis_3k_ion_corr}
\end{figure*}
Fig.~\ref{fig:psbis_3k_ion_corr} shows the ratios between the power spectra and bispectra of equilateral triangles without and with correction.

At $z \gtrsim 13$, the $\Delta_{\rm 21cm}$ without correction is underestimated e.g. at $z = 16$ by $\sim 30\%$ at $k = 0.3 \,\rm Mpc^{-1}$, $\sim 20\%$ at $k = 1 \,\rm Mpc^{-1}$, while only $\sim 10\%$ at $k = 3 \,\rm Mpc^{-1}$.
The ratios in the GAL and GALQSO models are smaller than in the other models, especially at $k = 3 \,\rm Mpc^{-1}$.
At $9 \lesssim z \lesssim 13$, the $\Delta_{\rm 21cm}$ without correction are higher than those after correction e.g. by $\sim 30\%$ at $z \sim 9$, with little dependence on $k$s.
The differences in the five models are only obvious at $k = 0.3 \,\rm Mpc^{-1}$, while they are small at $k = 1$ and $3 \,\rm Mpc^{-1}$.
As at $z<8$, the neutral gas is highly heated in the GALXRB, GALISM and GXQI models, the effects of correction on the 21 cm power spectra are negligible, i.e. the ratios are $\sim 1$, while in the GAL and GALQSO models the $\Delta_{\rm 21cm}$ without correction are $> 20\%$ higher than those after correction.

The bispectra without correction are typically overestimated at $z>8$, depending on the models and $k$s, e.g. by more than $200\%$ in the GALISM, GALXRB and GXQI models and $\sim 100\%$ in the GAL and GALQSO models at $z=16$ and $k = 0.3 \,\rm Mpc^{-1}$, while by $\sim 30\%$ at $k = 1 \,\rm Mpc^{-1}$ and $\sim 50\%$ at $k = 3 \,\rm Mpc^{-1}$ at the same redshift.
At $z<8$, the bispectra without correction are close to those after correction in the GALISM, GALXRB and GXQI models, i.e. the ratios are $\sim 1$, while without correction the bispectra in the GAL and GALQSO models are overestimated, especially at $k = 3 \,\rm Mpc^{-1}$.
The exception of the GXQI model at $k = 0.3 \,\rm Mpc^{-1}$ is due to the numerical error on the bispectrum at large scales (see also Appendix A).
Finally, we note that without correction the sign of bispectra is not obviously changed.
}
\bibliographystyle{aasjournal}
\bibliography{ref}

\begin{thebibliography}{}
\expandafter\ifx\csname natexlab\endcsname\relax\def\natexlab#1{#1}\fi
\providecommand{\url}[1]{\href{#1}{#1}}
\providecommand{\dodoi}[1]{doi:~\href{http://doi.org/#1}{\nolinkurl{#1}}}
\providecommand{\doeprint}[1]{\href{http://ascl.net/#1}{\nolinkurl{http://ascl.net/#1}}}
\providecommand{\doarXiv}[1]{\href{https://arxiv.org/abs/#1}{\nolinkurl{https://arxiv.org/abs/#1}}}

\bibitem[{{Alvarez}(2015)}]{Alvarez2015}
{Alvarez}, M.~A. 2015, ArXiv e-prints.
\newblock \doarXiv{1511.02846}

\bibitem[{{Baek} {et~al.}(2010){Baek}, {Semelin}, {Di Matteo}, {Revaz}, \&
  {Combes}}]{Baek2010}
{Baek}, S., {Semelin}, B., {Di Matteo}, P., {Revaz}, Y., \& {Combes}, F. 2010,
  \aap, 523, A4, \dodoi{10.1051/0004-6361/201014347}

\bibitem[{{Beane} \& {Lidz}(2018)}]{Beane2018}
{Beane}, A., \& {Lidz}, A. 2018, \apj, 867, 26,
  \dodoi{10.3847/1538-4357/aae388}

\bibitem[{{Bharadwaj} {et~al.}(2020){Bharadwaj}, {Mazumdar}, \&
  {Sarkar}}]{Bharadwaj2020}
{Bharadwaj}, S., {Mazumdar}, A., \& {Sarkar}, D. 2020, \mnras, 493, 594,
  \dodoi{10.1093/mnras/staa279}

\bibitem[{{Bhatt} {et~al.}(2020){Bhatt}, {Natwariya}, {Nayak}, \&
  {Pandey}}]{Bhatt2020}
{Bhatt}, J.~R., {Natwariya}, P.~K., {Nayak}, A.~C., \& {Pandey}, A.~K. 2020,
  European Physical Journal C, 80, 334, \dodoi{10.1140/epjc/s10052-020-7886-x}

\bibitem[{{Bowman} {et~al.}(2018){Bowman}, {Rogers}, {Monsalve}, {Mozdzen}, \&
  {Mahesh}}]{Bowman2018}
{Bowman}, J.~D., {Rogers}, A. E.~E., {Monsalve}, R.~A., {Mozdzen}, T.~J., \&
  {Mahesh}, N. 2018, \nat, 555, 67, \dodoi{10.1038/nature25792}

\bibitem[{{Busch} {et~al.}(2020){Busch}, {Eide}, {Ciardi}, \&
  {Kakiichi}}]{Busch2020}
{Busch}, P., {Eide}, M.~B., {Ciardi}, B., \& {Kakiichi}, K. 2020, \mnras, 498,
  4533, \dodoi{10.1093/mnras/staa2599}

\bibitem[{{Choudhury} {et~al.}(2015){Choudhury}, {Puchwein}, {Haehnelt}, \&
  {Bolton}}]{Choudhury2015}
{Choudhury}, T.~R., {Puchwein}, E., {Haehnelt}, M.~G., \& {Bolton}, J.~S. 2015,
  \mnras, 452, 261, \dodoi{10.1093/mnras/stv1250}

\bibitem[{{Christian} \& {Loeb}(2013)}]{Christian2013}
{Christian}, P., \& {Loeb}, A. 2013, \jcap, 9, 014,
  \dodoi{10.1088/1475-7516/2013/09/014}

\bibitem[{{Ciardi} {et~al.}(2001){Ciardi}, {Ferrara}, {Marri}, \&
  {Raimondo}}]{Ciardi2001}
{Ciardi}, B., {Ferrara}, A., {Marri}, S., \& {Raimondo}, G. 2001, \mnras, 324,
  381, \dodoi{10.1046/j.1365-8711.2001.04316.x}

\bibitem[{{Ciardi} \& {Madau}(2003)}]{Ciardi2003}
{Ciardi}, B., \& {Madau}, P. 2003, \apj, 596, 1, \dodoi{10.1086/377634}

\bibitem[{{Cohen} {et~al.}(2018){Cohen}, {Fialkov}, \& {Barkana}}]{Cohen2018}
{Cohen}, A., {Fialkov}, A., \& {Barkana}, R. 2018, \mnras, 478, 2193,
  \dodoi{10.1093/mnras/sty1094}

\bibitem[{{Dewdney et al.}(2016)}]{Dewdney2016}
{Dewdney et al.} 2016, {SKA1 SYSTEM BASELINE DESIGN V2},
  \url{https://astronomers.skatelescope.org/wp-content/uploads/2016/05/SKA-TEL-SKO-0000002_03_SKA1SystemBaselineDesignV2.pdf}

\bibitem[{{Eide} {et~al.}(2020{\natexlab{a}}){Eide}, {Ciardi}, {Feng}, \& {Di
  Matteo}}]{Eide2020b}
{Eide}, M.~B., {Ciardi}, B., {Feng}, Y., \& {Di Matteo}, T. 2020{\natexlab{a}},
  \mnras, \dodoi{10.1093/mnras/staa3253}

\bibitem[{{Eide} {et~al.}(2020{\natexlab{b}}){Eide}, {Ciardi}, {Graziani},
  {Busch}, {Feng}, \& {Matteo}}]{Eide2020}
{Eide}, M.~B., {Ciardi}, B., {Graziani}, L., {et~al.} 2020{\natexlab{b}},
  \mnras, \dodoi{10.1093/mnras/staa2774}

\bibitem[{{Eide} {et~al.}(2018){Eide}, {Graziani}, {Ciardi}, {Feng},
  {Kakiichi}, \& {Di Matteo}}]{Eide2018}
{Eide}, M.~B., {Graziani}, L., {Ciardi}, B., {et~al.} 2018, \mnras,
  \dodoi{10.1093/mnras/sty272}

\bibitem[{{Eldridge} \& {Stanway}(2012)}]{Eldridge2012}
{Eldridge}, J.~J., \& {Stanway}, E.~R. 2012, \mnras, 419, 479,
  \dodoi{10.1111/j.1365-2966.2011.19713.x}

\bibitem[{{Fan} {et~al.}(2006){Fan}, {Carilli}, \& {Keating}}]{Fan2006}
{Fan}, X., {Carilli}, C.~L., \& {Keating}, B. 2006, \araa, 44, 415,
  \dodoi{10.1146/annurev.astro.44.051905.092514}

\bibitem[{{Fan} {et~al.}(2003){Fan}, {Strauss}, {Schneider}, {Becker}, {White},
  {Haiman}, {Gregg}, {Pentericci}, {Grebel}, {Narayanan}, {Loh}, {Richards},
  {Gunn}, {Lupton}, {Knapp}, {Ivezi{\'c}}, {Brandt}, {Collinge}, {Hao},
  {Harbeck}, {Prada}, {Schaye}, {Strateva}, {Zakamska}, {Anderson},
  {Brinkmann}, {Bahcall}, {Lamb}, {Okamura}, {Szalay}, \& {York}}]{Fan2003}
{Fan}, X., {Strauss}, M.~A., {Schneider}, D.~P., {et~al.} 2003, \aj, 125, 1649,
  \dodoi{10.1086/368246}

\bibitem[{{Fialkov} {et~al.}(2014){Fialkov}, {Barkana}, \&
  {Visbal}}]{Fialkov2014}
{Fialkov}, A., {Barkana}, R., \& {Visbal}, E. 2014, \nat, 506, 197,
  \dodoi{10.1038/nature12999}

\bibitem[{{Fialkov} {et~al.}(2017){Fialkov}, {Cohen}, {Barkana}, \&
  {Silk}}]{Fialkov2017}
{Fialkov}, A., {Cohen}, A., {Barkana}, R., \& {Silk}, J. 2017, \mnras, 464,
  3498, \dodoi{10.1093/mnras/stw2540}

\bibitem[{{Field}(1959)}]{Field1959}
{Field}, G.~B. 1959, \apj, 129, 536, \dodoi{10.1086/146653}

\bibitem[{{Fragos} {et~al.}(2013{\natexlab{a}}){Fragos}, {Lehmer}, {Naoz},
  {Zezas}, \& {Basu-Zych}}]{Fragos2013b}
{Fragos}, T., {Lehmer}, B.~D., {Naoz}, S., {Zezas}, A., \& {Basu-Zych}, A.
  2013{\natexlab{a}}, \apjl, 776, L31, \dodoi{10.1088/2041-8205/776/2/L31}

\bibitem[{{Fragos} {et~al.}(2013{\natexlab{b}}){Fragos}, {Lehmer}, {Tremmel},
  {Tzanavaris}, {Basu-Zych}, {Belczynski}, {Hornschemeier}, {Jenkins},
  {Kalogera}, {Ptak}, \& {Zezas}}]{Fragos2013a}
{Fragos}, T., {Lehmer}, B., {Tremmel}, M., {et~al.} 2013{\natexlab{b}}, \apj,
  764, 41, \dodoi{10.1088/0004-637X/764/1/41}

\bibitem[{{Furlanetto} {et~al.}(2006){Furlanetto}, {Oh}, \&
  {Briggs}}]{Furlanetto2006}
{Furlanetto}, S.~R., {Oh}, S.~P., \& {Briggs}, F.~H. 2006, \physrep, 433, 181,
  \dodoi{10.1016/j.physrep.2006.08.002}

\bibitem[{{Geil} \& {Wyithe}(2009)}]{Geil2009}
{Geil}, P.~M., \& {Wyithe}, J.~S.~B. 2009, \mnras, 399, 1877,
  \dodoi{10.1111/j.1365-2966.2009.15451.x}

\bibitem[{{Geil} {et~al.}(2008){Geil}, {Wyithe}, {Petrovic}, \&
  {Oh}}]{Geil2008}
{Geil}, P.~M., {Wyithe}, J.~S.~B., {Petrovic}, N., \& {Oh}, S.~P. 2008, \mnras,
  390, 1496, \dodoi{10.1111/j.1365-2966.2008.13798.x}

\bibitem[{{Ghara} {et~al.}(2020){Ghara}, {Giri}, {Mellema}, {Ciardi},
  {Zaroubi}, {Iliev}, {Koopmans}, {Chapman}, {Gazagnes}, {Gehlot}, {Ghosh},
  {Jeli{\'c}}, {Mertens}, {Mondal}, {Schaye}, {Silva}, {Asad}, {Kooistra},
  {Mevius}, {Offringa}, {Pandey}, \& {Yatawatta}}]{Ghara2020}
{Ghara}, R., {Giri}, S.~K., {Mellema}, G., {et~al.} 2020, \mnras, 493, 4728,
  \dodoi{10.1093/mnras/staa487}

\bibitem[{{Gillet} {et~al.}(2018){Gillet}, {Mesinger}, {Greig}, {Liu}, \&
  {Ucci}}]{Gillet2018}
{Gillet}, N., {Mesinger}, A., {Greig}, B., {Liu}, A., \& {Ucci}, G. 2018, arXiv
  e-prints.
\newblock \doarXiv{1805.02699}

\bibitem[{{Giri} {et~al.}(2019){Giri}, {D'Aloisio}, {Mellema}, {Komatsu},
  {Ghara}, \& {Majumdar}}]{Giri2019}
{Giri}, S.~K., {D'Aloisio}, A., {Mellema}, G., {et~al.} 2019, \jcap, 2019, 058,
  \dodoi{10.1088/1475-7516/2019/02/058}

\bibitem[{{Glatzle} {et~al.}(2019){Glatzle}, {Ciardi}, \&
  {Graziani}}]{Glatzle2019}
{Glatzle}, M., {Ciardi}, B., \& {Graziani}, L. 2019, \mnras, 482, 321,
  \dodoi{10.1093/mnras/sty2514}

\bibitem[{{Graziani} {et~al.}(2018){Graziani}, {Ciardi}, \&
  {Glatzle}}]{Graziani2018}
{Graziani}, L., {Ciardi}, B., \& {Glatzle}, M. 2018, \mnras, 479, 4320,
  \dodoi{10.1093/mnras/sty1367}

\bibitem[{{Graziani} {et~al.}(2013){Graziani}, {Maselli}, \&
  {Ciardi}}]{Graziani2013}
{Graziani}, L., {Maselli}, A., \& {Ciardi}, B. 2013, \mnras, 431, 722,
  \dodoi{10.1093/mnras/stt206}

\bibitem[{{Greig} {et~al.}(2020){Greig}, {Trott}, {Barry}, {Mutch}, {Pindor},
  {Webster}, \& {Wyithe}}]{Greig2020}
{Greig}, B., {Trott}, C.~M., {Barry}, N., {et~al.} 2020, arXiv e-prints,
  arXiv:2008.02639.
\newblock \doarXiv{2008.02639}

\bibitem[{{Harker} {et~al.}(2010){Harker}, {Zaroubi}, {Bernardi}, {Brentjens},
  {de Bruyn}, {Ciardi}, {Jeli{\'c}}, {Koopmans}, {Labropoulos}, {Mellema},
  {Offringa}, {Pandey}, {Pawlik}, {Schaye}, {Thomas}, \&
  {Yatawatta}}]{Harker2010}
{Harker}, G., {Zaroubi}, S., {Bernardi}, G., {et~al.} 2010, \mnras, 405, 2492,
  \dodoi{10.1111/j.1365-2966.2010.16628.x}

\bibitem[{{Hassan} {et~al.}(2020){Hassan}, {Andrianomena}, \&
  {Doughty}}]{Hassan2020}
{Hassan}, S., {Andrianomena}, S., \& {Doughty}, C. 2020, \mnras, 494, 5761,
  \dodoi{10.1093/mnras/staa1151}

\bibitem[{{Hektor} {et~al.}(2018){Hektor}, {H{\"u}tsi}, {Marzola}, {Raidal},
  {Vaskonen}, \& {Veerm{\"a}e}}]{Hektor2018}
{Hektor}, A., {H{\"u}tsi}, G., {Marzola}, L., {et~al.} 2018, \prd, 98, 023503,
  \dodoi{10.1103/PhysRevD.98.023503}

\bibitem[{{Hills} {et~al.}(2018){Hills}, {Kulkarni}, {Meerburg}, \&
  {Puchwein}}]{Hills2018}
{Hills}, R., {Kulkarni}, G., {Meerburg}, P.~D., \& {Puchwein}, E. 2018, \nat,
  564, E32, \dodoi{10.1038/s41586-018-0796-5}

\bibitem[{{Hoffmann} {et~al.}(2019){Hoffmann}, {Mao}, {Xu}, {Mo}, \& {Wand
  elt}}]{Hoffmann2019}
{Hoffmann}, K., {Mao}, Y., {Xu}, J., {Mo}, H., \& {Wand elt}, B.~D. 2019,
  \mnras, 487, 3050, \dodoi{10.1093/mnras/stz1472}

\bibitem[{{Hutter} {et~al.}(2020){Hutter}, {Watkinson}, {Seiler}, {Dayal},
  {Sinha}, \& {Croton}}]{Hutter2020}
{Hutter}, A., {Watkinson}, C.~A., {Seiler}, J., {et~al.} 2020, \mnras, 492,
  653, \dodoi{10.1093/mnras/stz3139}

\bibitem[{{Jeli{\'c}} {et~al.}(2010){Jeli{\'c}}, {Zaroubi}, {Aghanim},
  {Douspis}, {Koopmans}, {Langer}, {Mellema}, {Tashiro}, \&
  {Thomas}}]{Jelic2010}
{Jeli{\'c}}, V., {Zaroubi}, S., {Aghanim}, N., {et~al.} 2010, \mnras, 402,
  2279, \dodoi{10.1111/j.1365-2966.2009.16086.x}

\bibitem[{{Khandai} {et~al.}(2015){Khandai}, {Di Matteo}, {Croft}, {Wilkins},
  {Feng}, {Tucker}, {DeGraf}, \& {Liu}}]{Khandai2015}
{Khandai}, N., {Di Matteo}, T., {Croft}, R., {et~al.} 2015, \mnras, 450, 1349,
  \dodoi{10.1093/mnras/stv627}

\bibitem[{{Khullar} {et~al.}(2020){Khullar}, {Ma}, {Busch}, {Ciardi}, {Eide},
  \& {Kakiichi}}]{Khullar2020}
{Khullar}, S., {Ma}, Q., {Busch}, P., {et~al.} 2020, \mnras, 497, 572,
  \dodoi{10.1093/mnras/staa1951}

\bibitem[{{Komatsu} {et~al.}(2011){Komatsu}, {Smith}, {Dunkley}, {Bennett},
  {Gold}, {Hinshaw}, {Jarosik}, {Larson}, {Nolta}, {Page}, {Spergel},
  {Halpern}, {Hill}, {Kogut}, {Limon}, {Meyer}, {Odegard}, {Tucker}, {Weiland},
  {Wollack}, \& {Wright}}]{Komatsu2011}
{Komatsu}, E., {Smith}, K.~M., {Dunkley}, J., {et~al.} 2011, \apjs, 192, 18,
  \dodoi{10.1088/0067-0049/192/2/18}

\bibitem[{{Koopmans} {et~al.}(2015){Koopmans}, {Pritchard}, {Mellema},
  {Aguirre}, {Ahn}, {Barkana}, {van Bemmel}, {Bernardi}, {Bonaldi}, {Briggs},
  {de Bruyn}, {Chang}, {Chapman}, {Chen}, {Ciardi}, {Dayal}, {Ferrara},
  {Fialkov}, {Fiore}, {Ichiki}, {Illiev}, {Inoue}, {Jelic}, {Jones}, {Lazio},
  {Maio}, {Majumdar}, {Mack}, {Mesinger}, {Morales}, {Parsons}, {Pen},
  {Santos}, {Schneider}, {Semelin}, {de Souza}, {Subrahmanyan}, {Takeuchi},
  {Vedantham}, {Wagg}, {Webster}, {Wyithe}, {Datta}, \& {Trott}}]{Koopmans2015}
{Koopmans}, L., {Pritchard}, J., {Mellema}, G., {et~al.} 2015, Advancing
  Astrophysics with the Square Kilometre Array (AASKA14), 1.
\newblock \doarXiv{1505.07568}

\bibitem[{{Kovetz} {et~al.}(2018){Kovetz}, {Poulin}, {Gluscevic}, {Boddy},
  {Barkana}, \& {Kamionkowski}}]{Kovetz2018}
{Kovetz}, E.~D., {Poulin}, V., {Gluscevic}, V., {et~al.} 2018, \prd, 98,
  103529, \dodoi{10.1103/PhysRevD.98.103529}

\bibitem[{{Krawczyk} {et~al.}(2013){Krawczyk}, {Richards}, {Mehta}, {Vogeley},
  {Gallagher}, {Leighly}, {Ross}, \& {Schneider}}]{Krawczyk2013}
{Krawczyk}, C.~M., {Richards}, G.~T., {Mehta}, S.~S., {et~al.} 2013, \apjs,
  206, 4, \dodoi{10.1088/0067-0049/206/1/4}

\bibitem[{{Kubota} {et~al.}(2016){Kubota}, {Yoshiura}, {Shimabukuro}, \&
  {Takahashi}}]{Kubota2016}
{Kubota}, K., {Yoshiura}, S., {Shimabukuro}, H., \& {Takahashi}, K. 2016,
  \pasj, 68, 61, \dodoi{10.1093/pasj/psw059}

\bibitem[{{La Plante} {et~al.}(2020){La Plante}, {Lidz}, {Aguirre}, \&
  {Kohn}}]{Plante2020}
{La Plante}, P., {Lidz}, A., {Aguirre}, J., \& {Kohn}, S. 2020, arXiv e-prints,
  arXiv:2005.07206.
\newblock \doarXiv{2005.07206}

\bibitem[{{Liang} {et~al.}(2016){Liang}, {Mao}, \& {Qin}}]{Liang2016}
{Liang}, J.-M., {Mao}, X.-C., \& {Qin}, B. 2016, Research in Astronomy and
  Astrophysics, 16, 132, \dodoi{10.1088/1674-4527/16/8/132}

\bibitem[{{Lidz} {et~al.}(2009){Lidz}, {Zahn}, {Furlanetto}, {McQuinn},
  {Hernquist}, \& {Zaldarriaga}}]{Lidz2009}
{Lidz}, A., {Zahn}, O., {Furlanetto}, S.~R., {et~al.} 2009, \apj, 690, 252,
  \dodoi{10.1088/0004-637X/690/1/252}

\bibitem[{{Ma} {et~al.}(2018{\natexlab{a}}){Ma}, {Ciardi}, {Eide}, \&
  {Helgason}}]{Ma2018b}
{Ma}, Q., {Ciardi}, B., {Eide}, M.~B., \& {Helgason}, K. 2018{\natexlab{a}},
  \mnras, 480, 26, \dodoi{10.1093/mnras/sty1806}

\bibitem[{{Ma} {et~al.}(2018{\natexlab{b}}){Ma}, {Helgason}, {Komatsu},
  {Ciardi}, \& {Ferrara}}]{Ma2018a}
{Ma}, Q., {Helgason}, K., {Komatsu}, E., {Ciardi}, B., \& {Ferrara}, A.
  2018{\natexlab{b}}, \mnras, 476, 4025, \dodoi{10.1093/mnras/sty543}

\bibitem[{{Ma} {et~al.}(2020){Ma}, {Ciardi}, {Kakiichi}, {Zaroubi}, {Zhi}, \&
  {Busch}}]{Ma2020}
{Ma}, Q.-B., {Ciardi}, B., {Kakiichi}, K., {et~al.} 2020, \apj, 888, 112,
  \dodoi{10.3847/1538-4357/ab5b95}

\bibitem[{{Madau} \& {Fragos}(2017)}]{Madau2017}
{Madau}, P., \& {Fragos}, T. 2017, \apj, 840, 39,
  \dodoi{10.3847/1538-4357/aa6af9}

\bibitem[{{Madau} {et~al.}(1997){Madau}, {Meiksin}, \& {Rees}}]{Madau1997}
{Madau}, P., {Meiksin}, A., \& {Rees}, M.~J. 1997, \apj, 475, 429

\bibitem[{{Majumdar} {et~al.}(2013){Majumdar}, {Bharadwaj}, \&
  {Choudhury}}]{Majumdar2013}
{Majumdar}, S., {Bharadwaj}, S., \& {Choudhury}, T.~R. 2013, \mnras, 434, 1978,
  \dodoi{10.1093/mnras/stt1144}

\bibitem[{{Majumdar} {et~al.}(2020){Majumdar}, {Kamran}, {Pritchard}, {Mondal},
  {Mazumdar}, {Bharadwaj}, \& {Mellema}}]{Majumdar2020}
{Majumdar}, S., {Kamran}, M., {Pritchard}, J.~R., {et~al.} 2020, arXiv
  e-prints, arXiv:2007.06584.
\newblock \doarXiv{2007.06584}

\bibitem[{{Majumdar} {et~al.}(2018){Majumdar}, {Pritchard}, {Mondal},
  {Watkinson}, {Bharadwaj}, \& {Mellema}}]{Majumdar2018}
{Majumdar}, S., {Pritchard}, J.~R., {Mondal}, R., {et~al.} 2018, \mnras, 476,
  4007, \dodoi{10.1093/mnras/sty535}

\bibitem[{{Majumdar} {et~al.}(2016){Majumdar}, {Jensen}, {Mellema}, {Chapman},
  {Abdalla}, {Lee}, {Iliev}, {Dixon}, {Datta}, {Ciardi}, {Fernandez},
  {Jeli{\'c}}, {Koopmans}, \& {Zaroubi}}]{Majumdar2016}
{Majumdar}, S., {Jensen}, H., {Mellema}, G., {et~al.} 2016, \mnras, 456, 2080,
  \dodoi{10.1093/mnras/stv2812}

\bibitem[{{Mangena} {et~al.}(2020){Mangena}, {Hassan}, \&
  {Santos}}]{Mangena2020}
{Mangena}, T., {Hassan}, S., \& {Santos}, M.~G. 2020, \mnras, 494, 600,
  \dodoi{10.1093/mnras/staa750}

\bibitem[{{Mao} {et~al.}(2012){Mao}, {Shapiro}, {Mellema}, {Iliev}, {Koda}, \&
  {Ahn}}]{Mao2012}
{Mao}, Y., {Shapiro}, P.~R., {Mellema}, G., {et~al.} 2012, \mnras, 422, 926,
  \dodoi{10.1111/j.1365-2966.2012.20471.x}

\bibitem[{{Maselli} {et~al.}(2009){Maselli}, {Ciardi}, \&
  {Kanekar}}]{Maselli2009}
{Maselli}, A., {Ciardi}, B., \& {Kanekar}, A. 2009, \mnras, 393, 171,
  \dodoi{10.1111/j.1365-2966.2008.14197.x}

\bibitem[{{Meerburg} {et~al.}(2013){Meerburg}, {Dvorkin}, \&
  {Spergel}}]{Meerburg2013}
{Meerburg}, P.~D., {Dvorkin}, C., \& {Spergel}, D.~N. 2013, \apj, 779, 124,
  \dodoi{10.1088/0004-637X/779/2/124}

\bibitem[{{Mertens} {et~al.}(2020){Mertens}, {Mevius}, {Koopmans}, {Offringa},
  {Mellema}, {Zaroubi}, {Brentjens}, {Gan}, {Gehlot}, {Pand ey}, {Sardarabadi},
  {Vedantham}, {Yatawatta}, {Asad}, {Ciardi}, {Chapman}, {Gazagnes}, {Ghara},
  {Ghosh}, {Giri}, {Iliev}, {Jeli{\'c}}, {Kooistra}, {Mondal}, {Schaye}, \&
  {Silva}}]{Mertens2020}
{Mertens}, F.~G., {Mevius}, M., {Koopmans}, L.~V.~E., {et~al.} 2020, \mnras,
  493, 1662, \dodoi{10.1093/mnras/staa327}

\bibitem[{{Mesinger} {et~al.}(2013){Mesinger}, {Ferrara}, \&
  {Spiegel}}]{Mesinger2013}
{Mesinger}, A., {Ferrara}, A., \& {Spiegel}, D.~S. 2013, \mnras, 431, 621,
  \dodoi{10.1093/mnras/stt198}

\bibitem[{{Mesinger} {et~al.}(2011){Mesinger}, {Furlanetto}, \&
  {Cen}}]{Mesinger2011}
{Mesinger}, A., {Furlanetto}, S., \& {Cen}, R. 2011, \mnras, 411, 955,
  \dodoi{10.1111/j.1365-2966.2010.17731.x}

\bibitem[{{Mineo} {et~al.}(2012){Mineo}, {Gilfanov}, \&
  {Sunyaev}}]{Mineo2012_ism}
{Mineo}, S., {Gilfanov}, M., \& {Sunyaev}, R. 2012, \mnras, 426, 1870,
  \dodoi{10.1111/j.1365-2966.2012.21831.x}

\bibitem[{{Mondal} {et~al.}(2015){Mondal}, {Bharadwaj}, {Majumdar}, {Bera}, \&
  {Acharyya}}]{Mondal2015}
{Mondal}, R., {Bharadwaj}, S., {Majumdar}, S., {Bera}, A., \& {Acharyya}, A.
  2015, \mnras, 449, L41, \dodoi{10.1093/mnrasl/slv015}

\bibitem[{{Mondal} {et~al.}(2020){Mondal}, {Fialkov}, {Fling}, {Iliev},
  {Barkana}, {Ciardi}, {Mellema}, {Zaroubi}, {Koopmans}, {Mertens}, {Gehlot},
  {Ghara}, {Ghosh}, {Giri}, {Offringa}, \& {Pand ey}}]{Mondal2020}
{Mondal}, R., {Fialkov}, A., {Fling}, C., {et~al.} 2020, \mnras,
  \dodoi{10.1093/mnras/staa2422}

\bibitem[{{Morales} \& {Wyithe}(2010)}]{Morales2010}
{Morales}, M.~F., \& {Wyithe}, J. S.~B. 2010, \araa, 48, 127,
  \dodoi{10.1146/annurev-astro-081309-130936}

\bibitem[{{Moriwaki} {et~al.}(2019){Moriwaki}, {Yoshida}, {Eide}, \&
  {Ciardi}}]{Moriwaki2019}
{Moriwaki}, K., {Yoshida}, N., {Eide}, M.~B., \& {Ciardi}, B. 2019, arXiv
  e-prints.
\newblock \doarXiv{1906.10863}

\bibitem[{{Ota} {et~al.}(2017){Ota}, {Iye}, {Kashikawa}, {Konno}, {Nakata},
  {Totani}, {Kobayashi}, {Fudamoto}, {Seko}, {Toshikawa}, {Ichikawa},
  {Shibuya}, \& {Onoue}}]{Ota2017}
{Ota}, K., {Iye}, M., {Kashikawa}, N., {et~al.} 2017, \apj, 844, 85,
  \dodoi{10.3847/1538-4357/aa7a0a}

\bibitem[{{Pacucci} {et~al.}(2014){Pacucci}, {Mesinger}, {Mineo}, \&
  {Ferrara}}]{Pacucci2014}
{Pacucci}, F., {Mesinger}, A., {Mineo}, S., \& {Ferrara}, A. 2014, \mnras, 443,
  678, \dodoi{10.1093/mnras/stu1240}

\bibitem[{{Patil} {et~al.}(2014){Patil}, {Zaroubi}, {Chapman}, {Jeli{\'c}},
  {Harker}, {Abdalla}, {Asad}, {Bernardi}, {Brentjens}, {de Bruyn}, {Bus},
  {Ciardi}, {Daiboo}, {Fernandez}, {Ghosh}, {Jensen}, {Kazemi}, {Koopmans},
  {Labropoulos}, {Mevius}, {Martinez}, {Mellema}, {Offringa}, {Pand ey},
  {Schaye}, {Thomas}, {Vedantham}, {Veligatla}, {Wijnholds}, \&
  {Yatawatta}}]{Patil2014}
{Patil}, A.~H., {Zaroubi}, S., {Chapman}, E., {et~al.} 2014, \mnras, 443, 1113,
  \dodoi{10.1093/mnras/stu1178}

\bibitem[{{Planck Collaboration} {et~al.}(2018){Planck Collaboration},
  {Aghanim}, {Akrami}, {Ashdown}, {Aumont}, {Baccigalupi}, {Ballardini},
  {Banday}, {Barreiro}, {Bartolo}, {Basak}, {Battye}, {Benabed}, {Bernard},
  {Bersanelli}, {Bielewicz}, {Bock}, {Bond}, {Borrill}, {Bouchet}, {Boulanger},
  {Bucher}, {Burigana}, {Butler}, {Calabrese}, {Cardoso}, {Carron},
  {Challinor}, {Chiang}, {Chluba}, {Colombo}, {Combet}, {Contreras}, {Crill},
  {Cuttaia}, {de Bernardis}, {de Zotti}, {Delabrouille}, {Delouis}, {Di
  Valentino}, {Diego}, {Dor{\'e}}, {Douspis}, {Ducout}, {Dupac}, {Dusini},
  {Efstathiou}, {Elsner}, {En{\ss}lin}, {Eriksen}, {Fantaye}, {Farhang},
  {Fergusson}, {Fernandez-Cobos}, {Finelli}, {Forastieri}, {Frailis},
  {Fraisse}, {Franceschi}, {Frolov}, {Galeotta}, {Galli}, {Ganga},
  {G{\'e}nova-Santos}, {Gerbino}, {Ghosh}, {Gonz{\'a}lez-Nuevo}, {G{\'o}rski},
  {Gratton}, {Gruppuso}, {Gudmundsson}, {Hamann}, {Handley}, {Hansen},
  {Herranz}, {Hildebrandt}, {Hivon}, {Huang}, {Jaffe}, {Jones}, {Karakci},
  {Keih{\"a}nen}, {Keskitalo}, {Kiiveri}, {Kim}, {Kisner}, {Knox},
  {Krachmalnicoff}, {Kunz}, {Kurki-Suonio}, {Lagache}, {Lamarre}, {Lasenby},
  {Lattanzi}, {Lawrence}, {Le Jeune}, {Lemos}, {Lesgourgues}, {Levrier},
  {Lewis}, {Liguori}, {Lilje}, {Lilley}, {Lindholm}, {L{\'o}pez-Caniego},
  {Lubin}, {Ma}, {Mac{\'\i}as-P{\'e}rez}, {Maggio}, {Maino}, {Mandolesi},
  {Mangilli}, {Marcos-Caballero}, {Maris}, {Martin}, {Martinelli},
  {Mart{\'\i}nez-Gonz{\'a}lez}, {Matarrese}, {Mauri}, {McEwen}, {Meinhold},
  {Melchiorri}, {Mennella}, {Migliaccio}, {Millea}, {Mitra},
  {Miville-Desch{\^e}nes}, {Molinari}, {Montier}, {Morgante}, {Moss}, {Natoli},
  {N{\o}rgaard-Nielsen}, {Pagano}, {Paoletti}, {Partridge}, {Patanchon},
  {Peiris}, {Perrotta}, {Pettorino}, {Piacentini}, {Polastri}, {Polenta},
  {Puget}, {Rachen}, {Reinecke}, {Remazeilles}, {Renzi}, {Rocha}, {Rosset},
  {Roudier}, {Rubi{\~n}o-Mart{\'\i}n}, {Ruiz-Granados}, {Salvati}, {Sandri},
  {Savelainen}, {Scott}, {Shellard}, {Sirignano}, {Sirri}, {Spencer},
  {Sunyaev}, {Suur-Uski}, {Tauber}, {Tavagnacco}, {Tenti}, {Toffolatti},
  {Tomasi}, {Trombetti}, {Valenziano}, {Valiviita}, {Van Tent}, {Vibert},
  {Vielva}, {Villa}, {Vittorio}, {Wand elt}, {Wehus}, {White}, {White},
  {Zacchei}, \& {Zonca}}]{Planck2018}
{Planck Collaboration}, {Aghanim}, N., {Akrami}, Y., {et~al.} 2018, arXiv
  e-prints, arXiv:1807.06209.
\newblock \doarXiv{1807.06209}

\bibitem[{{Pritchard} \& {Furlanetto}(2007)}]{Pritchard2007}
{Pritchard}, J.~R., \& {Furlanetto}, S.~R. 2007, \mnras, 376, 1680,
  \dodoi{10.1111/j.1365-2966.2007.11519.x}

\bibitem[{Rosenfeld \& Pfaltz(1966)}]{Rosenfeld1966}
Rosenfeld, A., \& Pfaltz, J.~L. 1966, J. ACM, 13, 471,
  \dodoi{10.1145/321356.321357}

\bibitem[{{Ross} {et~al.}(2019){Ross}, {Dixon}, {Ghara}, {Iliev}, \&
  {Mellema}}]{Ross2019}
{Ross}, H.~E., {Dixon}, K.~L., {Ghara}, R., {Iliev}, I.~T., \& {Mellema}, G.
  2019, \mnras, 487, 1101, \dodoi{10.1093/mnras/stz1220}

\bibitem[{{Ross} {et~al.}(2017){Ross}, {Dixon}, {Iliev}, \&
  {Mellema}}]{Ross2017}
{Ross}, H.~E., {Dixon}, K.~L., {Iliev}, I.~T., \& {Mellema}, G. 2017, \mnras,
  468, 3785, \dodoi{10.1093/mnras/stx649}

\bibitem[{{Ross} {et~al.}(2020){Ross}, {Giri}, {Dixon}, {Ghara}, {Iliev}, \&
  {Mellema}}]{Ross2020}
{Ross}, H.~E., {Giri}, S.~K., {Dixon}, K.~L., {et~al.} 2020, arXiv e-prints,
  arXiv:2011.03558.
\newblock \doarXiv{2011.03558}

\bibitem[{{Roy} {et~al.}(2020){Roy}, {Lapi}, {Spergel}, {Basak}, \&
  {Baccigalupi}}]{Roy2020}
{Roy}, A., {Lapi}, A., {Spergel}, D., {Basak}, S., \& {Baccigalupi}, C. 2020,
  \jcap, 2020, 062, \dodoi{10.1088/1475-7516/2020/03/062}

\bibitem[{{Seiler} {et~al.}(2018){Seiler}, {Hutter}, {Sinha}, \&
  {Croton}}]{Seiler2018}
{Seiler}, J., {Hutter}, A., {Sinha}, M., \& {Croton}, D. 2018, \mnras, 480,
  L33, \dodoi{10.1093/mnrasl/sly122}

\bibitem[{{Shakura} \& {Sunyaev}(1973)}]{Shakura1973}
{Shakura}, N.~I., \& {Sunyaev}, R.~A. 1973, \aap, 24, 337

\bibitem[{{Shaw} {et~al.}(2020){Shaw}, {Bharadwaj}, \& {Mondal}}]{Shaw2020}
{Shaw}, A.~K., {Bharadwaj}, S., \& {Mondal}, R. 2020, \mnras,
  \dodoi{10.1093/mnras/staa2090}

\bibitem[{{Shimabukuro} {et~al.}(2016){Shimabukuro}, {Yoshiura}, {Takahashi},
  {Yokoyama}, \& {Ichiki}}]{Shimabukuro2016}
{Shimabukuro}, H., {Yoshiura}, S., {Takahashi}, K., {Yokoyama}, S., \&
  {Ichiki}, K. 2016, \mnras, 458, 3003, \dodoi{10.1093/mnras/stw482}

\bibitem[{{Shimabukuro} {et~al.}(2017){Shimabukuro}, {Yoshiura}, {Takahashi},
  {Yokoyama}, \& {Ichiki}}]{Shimabukuro2017}
---. 2017, \mnras, 468, 1542, \dodoi{10.1093/mnras/stx530}

\bibitem[{{Thyagarajan} {et~al.}(2020){Thyagarajan}, {Carilli}, {Nikolic},
  {Kent}, {Mesinger}, {Kern}, {Bernardi}, {Matika}, {Abdurashidova}, {Aguirre},
  {Alexander}, {Ali}, {Balfour}, {Beardsley}, {Billings}, {Bowman}, {Bradley},
  {Burba}, {Carey}, {Cheng}, {DeBoer}, {Dexter}, {Acedo}, {Dillon}, {Ely},
  {Ewall-Wice}, {Fagnoni}, {Fritz}, {Furlanetto}, {Gale-Sides}, {Glendenning},
  {Gorthi}, {Greig}, {Grobbelaar}, {Halday}, {Hazelton}, {Hewitt}, {Hickish},
  {Jacobs}, {Julius}, {Kerrigan}, {Kittiwisit}, {Kohn}, {Kolopanis}, {Lanman},
  {La Plante}, {Lekalake}, {Lewis}, {Liu}, {MacMahon}, {Malan}, {Malgas},
  {Maree}, {Martinot}, {Matsetela}, {Molewa}, {Morales}, {Mosiane}, {Neben},
  {Parsons}, {Patra}, {Pieterse}, {Pober}, {Razavi-Ghods}, {Ringuette},
  {Robnett}, {Rosie}, {Sims}, {Smith}, {Syce}, {Williams}, \&
  {Zheng}}]{Thyagarajan2020}
{Thyagarajan}, N., {Carilli}, C.~L., {Nikolic}, B., {et~al.} 2020, \prd, 102,
  022002, \dodoi{10.1103/PhysRevD.102.022002}

\bibitem[{{Tozzi} {et~al.}(2000){Tozzi}, {Madau}, {Meiksin}, \&
  {Rees}}]{Tozzi2000}
{Tozzi}, P., {Madau}, P., {Meiksin}, A., \& {Rees}, M.~J. 2000, \apj, 528, 597,
  \dodoi{10.1086/308196}

\bibitem[{{Trott} {et~al.}(2020){Trott}, {Jordan}, {Midgley}, {Barry}, {Greig},
  {Pindor}, {Cook}, {Sleap}, {Tingay}, {Ung}, {Hancock}, {Williams}, {Bowman},
  {Byrne}, {Chokshi}, {Hazelton}, {Hasegawa}, {Jacobs}, {Joseph}, {Li}, {Line},
  {Lynch}, {McKinley}, {Mitchell}, {Morales}, {Ouchi}, {Pober}, {Rahimi},
  {Takahashi}, {Wayth}, {Webster}, {Wilensky}, {Wyithe}, {Yoshiura}, {Zhang},
  \& {Zheng}}]{Trott2020}
{Trott}, C.~M., {Jordan}, C.~H., {Midgley}, S., {et~al.} 2020, \mnras, 493,
  4711, \dodoi{10.1093/mnras/staa414}

\bibitem[{{van Haarlem} {et~al.}(2013){van Haarlem}, {Wise}, {Gunst}, {Heald},
  {McKean}, {Hessels}, {de Bruyn}, {Nijboer}, {Swinbank}, {Fallows},
  {Brentjens}, {Nelles}, {Beck}, {Falcke}, {Fender}, {H{\"o}randel},
  {Koopmans}, {Mann}, {Miley}, {R{\"o}ttgering}, {Stappers}, {Wijers},
  {Zaroubi}, {van den Akker}, {Alexov}, {Anderson}, {Anderson}, {van Ardenne},
  {Arts}, {Asgekar}, {Avruch}, {Batejat}, {B{\"a}hren}, {Bell}, {Bell}, {van
  Bemmel}, {Bennema}, {Bentum}, {Bernardi}, {Best}, {B{\^i}rzan}, {Bonafede},
  {Boonstra}, {Braun}, {Bregman}, {Breitling}, {van de Brink}, {Broderick},
  {Broekema}, {Brouw}, {Br{\"u}ggen}, {Butcher}, {van Cappellen}, {Ciardi},
  {Coenen}, {Conway}, {Coolen}, {Corstanje}, {Damstra}, {Davies}, {Deller},
  {Dettmar}, {van Diepen}, {Dijkstra}, {Donker}, {Doorduin}, {Dromer}, {Drost},
  {van Duin}, {Eisl{\"o}ffel}, {van Enst}, {Ferrari}, {Frieswijk}, {Gankema},
  {Garrett}, {de Gasperin}, {Gerbers}, {de Geus}, {Grie{\ss}meier}, {Grit},
  {Gruppen}, {Hamaker}, {Hassall}, {Hoeft}, {Holties}, {Horneffer}, {van der
  Horst}, {van Houwelingen}, {Huijgen}, {Iacobelli}, {Intema}, {Jackson},
  {Jelic}, {de Jong}, {Juette}, {Kant}, {Karastergiou}, {Koers}, {Kollen},
  {Kondratiev}, {Kooistra}, {Koopman}, {Koster}, {Kuniyoshi}, {Kramer},
  {Kuper}, {Lambropoulos}, {Law}, {van Leeuwen}, {Lemaitre}, {Loose}, {Maat},
  {Macario}, {Markoff}, {Masters}, {McFadden}, {McKay-Bukowski}, {Meijering},
  {Meulman}, {Mevius}, {Middelberg}, {Millenaar}, {Miller-Jones}, {Mohan},
  {Mol}, {Morawietz}, {Morganti}, {Mulcahy}, {Mulder}, {Munk}, {Nieuwenhuis},
  {van Nieuwpoort}, {Noordam}, {Norden}, {Noutsos}, {Offringa}, {Olofsson},
  {Omar}, {Orr{\'u}}, {Overeem}, {Paas}, {Pandey-Pommier}, {Pandey}, {Pizzo},
  {Polatidis}, {Rafferty}, {Rawlings}, {Reich}, {de Reijer}, {Reitsma},
  {Renting}, {Riemers}, {Rol}, {Romein}, {Roosjen}, {Ruiter}, {Scaife}, {van
  der Schaaf}, {Scheers}, {Schellart}, {Schoenmakers}, {Schoonderbeek},
  {Serylak}, {Shulevski}, {Sluman}, {Smirnov}, {Sobey}, {Spreeuw}, {Steinmetz},
  {Sterks}, {Stiepel}, {Stuurwold}, {Tagger}, {Tang}, {Tasse}, {Thomas},
  {Thoudam}, {Toribio}, {van der Tol}, {Usov}, {van Veelen}, {van der Veen},
  {ter Veen}, {Verbiest}, {Vermeulen}, {Vermaas}, {Vocks}, {Vogt}, {de Vos},
  {van der Wal}, {van Weeren}, {Weggemans}, {Weltevrede}, {White}, {Wijnholds},
  {Wilhelmsson}, {Wucknitz}, {Yatawatta}, {Zarka}, {Zensus}, \& {van
  Zwieten}}]{Haarlem2013}
{van Haarlem}, M.~P., {Wise}, M.~W., {Gunst}, A.~W., {et~al.} 2013, \aap, 556,
  A2, \dodoi{10.1051/0004-6361/201220873}

\bibitem[{{Vrbanec} {et~al.}(2016){Vrbanec}, {Ciardi}, {Jeli{\'c}}, {Jensen},
  {Zaroubi}, {Fernandez}, {Ghosh}, {Iliev}, {Kakiichi}, {Koopmans}, \&
  {Mellema}}]{Vrbanec2016}
{Vrbanec}, D., {Ciardi}, B., {Jeli{\'c}}, V., {et~al.} 2016, \mnras, 457, 666,
  \dodoi{10.1093/mnras/stv2993}

\bibitem[{{Watkinson} {et~al.}(2019){Watkinson}, {Giri}, {Ross}, {Dixon},
  {Iliev}, {Mellema}, \& {Pritchard}}]{Watkinson2019}
{Watkinson}, C.~A., {Giri}, S.~K., {Ross}, H.~E., {et~al.} 2019, \mnras, 482,
  2653, \dodoi{10.1093/mnras/sty2740}

\bibitem[{{Watkinson} {et~al.}(2017){Watkinson}, {Majumdar}, {Pritchard}, \&
  {Mondal}}]{Watkinson2017}
{Watkinson}, C.~A., {Majumdar}, S., {Pritchard}, J.~R., \& {Mondal}, R. 2017,
  \mnras, 472, 2436, \dodoi{10.1093/mnras/stx2130}

\bibitem[{{Watkinson} {et~al.}(2020){Watkinson}, {Trott}, \&
  {Hothi}}]{Watkinson2020}
{Watkinson}, C.~A., {Trott}, C.~M., \& {Hothi}, I. 2020, arXiv e-prints,
  arXiv:2002.05992.
\newblock \doarXiv{2002.05992}

\bibitem[{{Weinberger} {et~al.}(2019){Weinberger}, {Haehnelt}, \&
  {Kulkarni}}]{Weinberger2019}
{Weinberger}, L.~H., {Haehnelt}, M.~G., \& {Kulkarni}, G. 2019, \mnras, 485,
  1350, \dodoi{10.1093/mnras/stz481}

\bibitem[{{Yoshiura} {et~al.}(2015){Yoshiura}, {Shimabukuro}, {Takahashi},
  {Momose}, {Nakanishi}, \& {Imai}}]{Yoshiura2015}
{Yoshiura}, S., {Shimabukuro}, H., {Takahashi}, K., {et~al.} 2015, \mnras, 451,
  266, \dodoi{10.1093/mnras/stv855}

\end{thebibliography}

\end{document}